\documentclass[manuscript]{aastex}

\usepackage{natbib}

\shorttitle{Granulation in Red Giants: Observations by {\it Kepler} and 3D Simulations}
\shortauthors{Mathur et al.}

\begin{document}


\title{Granulation in Red Giants: observations by the {\it Kepler mission} and 3D convection simulations}


\author{S. Mathur\altaffilmark{1},  S. Hekker\altaffilmark{2,3}, R. Trampedach\altaffilmark{4}, J. Ballot\altaffilmark{5,6}, T. Kallinger\altaffilmark{7,8}, D. Buzasi\altaffilmark{9}, R. A. Garc\'ia\altaffilmark{10}, D. Huber\altaffilmark{11},  A. Jim\'enez\altaffilmark{12,13}, B. Mosser\altaffilmark{14}, T.~R. Bedding\altaffilmark{11}, Y. Elsworth\altaffilmark{3},  C. R\'egulo\altaffilmark{12,13}, D. Stello\altaffilmark{11},  W.~J. Chaplin\altaffilmark{3}, J. De Ridder\altaffilmark{8}, S.~J. Hale\altaffilmark{3}, K. Kinemuchi\altaffilmark{15}, H. Kjeldsen\altaffilmark{16}, F. Mullally\altaffilmark{17}, S.~E. Thompson\altaffilmark{17}}
%\affil{NCAR, P.O. Box, Boulder, CO 80302}


\altaffiltext{1}{High Altitude Observatory, NCAR, P.O. Box 3000, Boulder, CO 80307, USA}
\altaffiltext{2}{Astronomical Institute "Anton Pannekoek", University of Amsterdam, PO Box 94249, 1090 GE Amsterdam, The Netherlands}
\altaffiltext{3}{School of Physics and Astronomy, University of Birmingham, Edgbaston, Birmingham B15 2TT, UK}
\altaffiltext{4}{JILA, University of Colorado, 440 UCB, Boulder, CO 80309, USA}
\altaffiltext{5}{Institut de Recherche en Astrophysique et Plan\'etologie, CNRS, 14 avenue E. Belin, 31400 Toulouse, France}
\altaffiltext{6}{Universit\'e de Toulouse, UPS-OMP, IRAP, 31400 Toulouse, France}
\altaffiltext{7}{Institute for Astronomy (IfA), University of Vienna, T\"urkenschanzstrasse 17, 1180 Vienna, Austria}
%\altaffiltext{8}{Department of Physics and Astronomy, University of British Columbia, 6224 Agricultural Road, Vancouver, BC V6T 1Z1, Canada}
\altaffiltext{8}{Instituut voor Sterrenkunde, K.U. Leuven, Celestijnenlaan 200D, 3001 Leuven, Belgium}
\altaffiltext{9}{Eureka Scientific, 2452 Delmer Street Suite 100, Oakland, CA 94602-3017, USA}
\altaffiltext{10}{Laboratoire AIM, CEA/DSM -- CNRS - Universit\'e Paris Diderot -- IRFU/SAp, 91191 Gif-sur-Yvette Cedex, France}
\altaffiltext{11}{Sydney Institute for Astronomy, School of Physics, University of Sydney, NSW 2006, Australia}
\altaffiltext{12}{Universidad de La Laguna, Dpto de Astrof\'isica, 38206, Tenerife, Spain}
\altaffiltext{13}{Instituto de Astrof\'\i sica de Canarias, 38205, La Laguna, Tenerife, Spain}
\altaffiltext{14}{LESIA, UMR8109, Universit\'e Pierre et Marie Curie, Universit\'e Denis Diderot, Obs. de Paris, 92195 Meudon Cedex, France}

\altaffiltext{15}{Bay Area Environmental Research Inst./NASA Ames Research Center, Moffett Field, CA 94035, USA}
\altaffiltext{16}{Danish AsteroSeismology Centre, Department of Physics and Astronomy, University of Aarhus, 8000 Aarhus C, Denmark}
\altaffiltext{17}{SETI Institute/NASA Ames Research Center, Moffett Field, CA 94035, USA}
%\altaffiltext{1}{Visiting Astronomer, Cerro Tololo Inter-American Observatory.
%CTIO is operated by AURA, Inc.\ under contract to the National Science
%\altaffiltext{2}{Society of Fellows, Harvard University.}
%\altaffiltext{3}{present address: Center for Astrophysics,
%    60 Garden Street, Cambridge, MA 02138}
%\altaffiltext{4}{Visiting Programmer, Space Telescope Science Institute}
%\altaffiltext{5}{Patron, Alonso's Bar and Grill}


\begin{abstract}

The granulation pattern that we observe on the surface of the Sun is due to hot plasma from the interior rising to the photosphere where it cools down, and descends back into the interior at the edges of granules. This is the visible manifestation of convection taking place in the outer part of the solar convection zone. Because red giants have deeper convection zones and more extended atmospheres than the Sun, we cannot a priori assume that granulation in red giants is a scaled version of solar granulation. Until now, neither observations nor 1D analytical convection models could put constraints on granulation in red giants. However, thanks to asteroseismology, this study can now be performed. The resulting parameters yield physical information about the granulation. We analyze $\sim$~1000 red giants that have been observed by {\it Kepler} during 13 months. We fit the power spectra with Harvey-like profiles to retrieve the characteristics of the granulation (time scale $\tau_{\rm gran}$ and power $P_{\rm gran}$). We also introduce a new time scale, $\tau_{\rm eff}$, which takes into account that different slopes are used in the Harvey functions. We search for a correlation between these parameters and the global acoustic-mode parameter (the position of maximum power, $\nu_{\rm max}$) as well as with stellar parameters (mass, radius, surface gravity (log $g$) and effective temperature ($T_{\rm eff}$)). We show that  $\tau_{\rm eff} \propto \nu_{\rm max}^{-0.89}$ and $P_{\rm gran} \propto \nu_{\rm max}^{-1.90}$, which is consistent with the theoretical predictions. We find that the granulation time scales of stars that belong to the red clump have similar values while the time scales of stars in the red-giant branch are spread in a wider range. Finally, we show that realistic 3D simulations of the surface convection in stars, spanning the ($T_{\rm eff}$, log $g$)-range of our sample of red giants, match the {\it Kepler} observations well in terms of trends.  

%{\bf Up to now, analytical models based on mixing-length theory fail to reproduce, for instance, temperature profiles, while radiative-hydrodynamical simulations of convection agree well with solar observations in terms of spectral line profiles.} 

%{\bf TBC.} Granulation: what we still do not understand. Simulations... need of more constraints. 
%We already have some scaling laws between $\nu_{\rm max}$, $\Delta \nu$, R, M, L, Teff...
%Study if there is a relation between acoustic parameters and granulation but also fundamental stellar parameters and granulations.
%Use different background fittings based on one or several Harvey laws.

\end{abstract}





\section{Introduction}

Granulation was first observed by \cite{1801RSPT...91..265H} on the Sun, and is widely known as the surface signature of convection, where bright cells of ascending hot gas are visible as the granules and the darker descending cool gas are the so-called ``intergranular lanes''. These solar granules have typical sizes of about 1\,Mm.
The study of granulation is tightly related to the analysis of convection
quantities, being the most important manifestation of convection at the
surface of the Sun in terms of energy. Other phenomena related to convection are the acoustic (p-mode) oscillations that reveal the internal structure of the Sun. These oscillations are stochastically excited in the convective atmosphere.

Red giants are cool, bright, and evolved stars. They have a surface gravity, log $g$, between 2 and 4 and an effective temperature, $T_{\rm eff}$, in the range $\sim$~4000 to $\sim$~6000\,K, as defined by \citet{2011AJ....141..108C}. For these stars, the mass is roughly in the range 0.7 to 4\,$M_{\odot}$ and log\,($L/L_{\odot}$) varies from 0.3 to 3 \citep{2009A&A...503L..21M,2011A&A...530A.100H}. They are interesting not only
because they provide constraints on distance, age, and chemical evolution of
stars, galaxies, and the extragalactic medium
\citep[e.g.][]{2001MNRAS.323..109G} but they also serve as
laboratories for studying convection, granulation and p-mode oscillations. As in the Sun, the inefficient, super-adiabatic
convection in the upper few pressure scale-heights is host to sonic,
turbulent flows that stochastically excite sound waves --- these stochastic
oscillations are also known as solar-like oscillations.
The unambiguous detection of non-radial
oscillations in red giants \citep{2009Natur.459..398D} from photometric time series
of several hundreds of red giants obtained with the CoRoT satellite
(Baglin et al. 2006) was a significant step forward in red-giant seismology.
These observations also allowed to study p-mode global parameters and their
scaling laws \citep{2009A&A...506..465H,2010A&A...517A..22M,2010AN....331..944M},
to estimate their masses and
radii \citep{2010A&A...509A..77K}, and to even retrieve evidence of
sharp features in their internal structure \citep{2010A&A...520L...6M} from, for instance,
the second ionization zone of helium. 

The launch of NASA's {\it Kepler Mission} \citep{2010Sci...327..977B} in March
2009, took the next big step in asteroseismology.  The {\it Kepler} field of view is 105 square degrees in the direction of Cygnus and Lyra and the filter of the photometer ranges from 4\,300 to 8\,800\,\AA, with a broad peak at 5\,900\,\AA. In total about 150\,000 stars are observed with high-precision photometry throughout the nominal lifetime (3.5 years) of the mission. While for $\sim$17\,000 red giants only a few months of data have been released to the public domain, \citep{2011AJ....141..108C,2011A&A...530A.100H}, around 1\,500 red
giants with time series longer than a year are at our disposal within the \textit{Kepler} Asteroseismic Science Consortium (KASC) 
\citep[][]{2010ApJ...713L.176B,2011A&A...525A.131H,2010ApJ...723.1607H,2010A&A...522A...1K,mosser2011}.
We thus have exquisite data for studying not only the  acoustic modes and their
 dependence on stellar parameters, but
we can also, for the first time, study properties of surface convection of
a large sample of red giants. The observations and data processing that
we have used are described in Section~\ref{sect:obs}.

When studying the p modes of stars, granulation is normally
considered a ``noise'' term and referred to as ``background". In the present
paper, however, we use this background signal to investigate the surface manifestation
of convection and the source of p-mode excitation.  In
Section~\ref{sect:granparm} different methods used to extract characteristic granulation parameters 
from the power spectra are described and compared. In Section~\ref{sect:scalings} we investigate the correlations between granulation properties, stellar parameters (mass, radius, surface gravity and $T_{\rm eff}$) and the frequency of maximum power, $\nu_{\rm max}$.

From  a modeling point of view, convection can be treated either by 1D
analytical models, or by 2 or 3D hydrodynamical simulations. 
The three most employed analytical models of
convection are: the mixing-length theory \citep[MLT;][]{bohm1958}, non-local
MLT formulations
\citep[e.g.][]{1967PASJ...19..140U,1977LNP....71...15G,1977ApJ...214..196G,2006ESASP.624E..78D},
and the Canuto and Mazzitelli model
\citep[CM;][]{1991ApJ...370..295C,1992ApJ...389..724C}.  All of these
formulations of convection are employed in stellar-structure modeling, but
have severe shortcomings in the surface layers, where most of the
approximations they are built on, break down \citep[for more details,
see][]{2010Ap&SS.328..213T}. In addition to these shortcomings, the free parameters of the analytical models suffer from large uncertainties.
  As a result, the analytical models have only limited predictive power.
 None of the analytical models deal with granulation 
  and estimates of sizes, contrasts and flow speeds rely on a
  number of additional assumptions.
  
  Contrary to 1D analytical models, realistic numerical simulations
  of convection are
  based on the quantum mechanics of the equation of state and opacities
  and the fundamental physics of radiative transfer and hydrodynamics.
 However, the simulations are also limited by the available computational
  power and hence employ various approximations to make them tractable.
Most of these approximations, such as the effect of a limited numerical resolution can be quantified \citep{2000A&A...359..669A}. 
  
 Although 2D simulations can be run at much higher resolution, the topology of convection is
fundamentally altered by the lack of a third dimension, and the properties of the 2D simulations
cannot be directly related to observations of stars. Since the top boundary of
convective envelopes occurs in the photosphere, a realistic treatment of
radiative transfer is also necessary when the aim is to produce simulations that can be
directly compared to observations \citep[][and references
therein]{2009LRSP....6....2N}. 
%Some 3D radiative-hydrodynamics simulations are also done for subgiants stars \citep{2004MNRAS.347.1208R}. 
The surface layers are, of course, the very
layers we observe, which is the reason why we turn to more realistic, 3D convection
simulations for guidance in the interpretation of the {\it Kepler} observations. 

Previously, \citet{2007A&A...469..687C} used 3D  red giant simulations for
  abundance analysis for [Fe/H] = 0 to -3, using the same code and atomic
  physics as presented here \citep{2000SoPh..192...91S}.  They find metallicity effects on the
  granulation with increased size of the granules for increased metallicity values. 
  Some 3D radiative-hydrodynamics simulations are also done for subgiant stars \citep{2004MNRAS.347.1208R}.
\cite{2009A&A...506..167L} compared their  3D radiative-hydrodynamics simulation carried out with CO$^5$BOLD code with the CoRoT observations of the F-type dwarf, HD~49933. They found a significant over-estimation of the theoretical signal by a factor of two to three in total power. 

 In this work, we use the simulations performed with the \citet{2000SoPh..192...91S} code and compare them to observations of $\sim$~1000 red giants.
  The simulations are introduced in Section~5, where we also explain the
synthetic {\it Kepler} data of these simulations and compare the results of the simulations with the observations. 

\section{Observations}
\label{sect:obs}

\subsection{Sample selection}
In the present investigation we used \textit{Kepler} observations of red giants showing solar-like oscillations, which were pre-selected for asteroseismic or astrometric purposes before launch of the spacecraft. While the asteroseismic sample is composed of subsamples with different selection criteria, the astrometric sample, which consists of the majority of our total sample, has been selected coherently. These $\sim$1\,000 stars are selected to be 1) as distant as possible, to ensure small parallaxes and proper motions, 2) as bright as possible, but with very small chances of saturating the detector in any season, 3) located in fields as uncrowded as possible, and 4) uniformly distributed over the focal plane and not close to the edge of a CCD. This resulted in the following criteria:
\begin{itemize}
\item $T_{\rm eff}$~$<$~5400~K
\item $\log g$~$<$~3.8 (c.g.s.)
\item 11.0~mag~$<$~\textit{Kepler} magnitude~$<$~12.5~mag
\item crowding~$>$~0.95 (i.e. very low probability of having light from another source)
\end{itemize}

$T_{\rm eff}$ and $\log g$ from the {\it Kepler Input Catalog} \citep[KIC,][]{2010ApJ...713L.109B, 2010ApJ...713L..79K,2011arXiv1102.0342B} have been used.

Long-cadence ($\sim$29.4 minute sampling) data of the first ten days (Q0), the initial one-month roll (Q1) and each of the following three-month rolls (Q2, Q3, Q4, and Q5) have been used to obtain long time series of more than one year of data  \citep{2010ApJ...713L..87J}. We have used data processed as described in \cite{2011MNRAS.414L...6G} in which instrumental effects, such as satellite safe-mode events or attitude adjustments, have been removed. These effects can affect the high-frequency range when jumps happen in the light curve or the low-frequency range when a safe-mode event is observed. Long-term variations have been preserved, which could contain signal of stellar origin, e.g., granulation, but also from instrumental effects. The latter are filtered out with a procedure described below (Section~2.2). 

Among the 1283 stars for which we have Q012345 data, we have selected a sample of $\sim$~1000 red giants (900 to 1100 depending on the method used to analyze them), corresponding to the ones where different teams returned values in agreement for the mean large separation, $\langle \Delta \nu \rangle$, and the position of maximum power, $\nu_{\rm max}$. See \cite{2011A&A...525A.131H} and references therein for descriptions and comparisons of the different methods and results.

\subsection{Removing instrumental long term variation}

From the long-term variations preserved in the data \citep{2011MNRAS.414L...6G}, we need to remove the instrumental signal without affecting the granulation signal. We assume that the characteristic granulation time scale $\tau_{\rm gran}$ is of the same order as the time scale of the oscillations, as is the case for the Sun \citep[see e.g.][]{2005A&A...443L..11V}. For red giants we therefore expect $\tau_{\rm gran}$ to be of the order of hours up to a few days for the largest stars of our sample. Note that these equivalent time scales differ by roughly a factor of $\pi$ in the frequency domain as an e-folding time of $\tau$ in power equates to a position in the frequency domain of $1/(\pi \tau)$, while a wave of period $T$ has a frequency of $1/T$.

We have investigated effects of different high-pass filters used to remove the signature of signals with periods longer than the expected $\tau_{\rm gran}$ for red giants, which we assume to be (partly) due to instrumental effects. We tested different filter shapes and decided to use a triangular smooth as it introduces less ripples due to sharp edges compared to boxcar filters. Additionally, we tested different filter widths of 10, 20, and 30 days. It showed that the wider filters result in increasingly different $\tau_{\rm gran}$ values obtained from the data, i.e. introducing a bias as well as additional scatter. From these results we concluded that a triangular smooth with a full width at half maximum of 10 days applied
to an interpolated time series provides most robust results. Note that the filter with a width of 10 days limits our sensitivity of $\tau_{\rm gran}$ to periods $<$\,10 days, which is sufficiently long to study the granulation in red giants under the aforementioned assumption that these are of the order of hours up to a few days.

%We have investigated different approaches to remove the signature of signals with periods longer than the expected $\tau_{\rm gran}$ for red giants, which we assume to be (partly) due to instrumental effects. This included tests with different shapes and widths of smoothing filters to compute a smoothed time series only containing the long-term variation, which we then removed from the data. From the tests it appeared that a triangular smooth, i.e., a double boxcar smooth of the time series, applied to an interpolated time series provides most robust results. For the full width at half maximum of the filter we currently use 10 days. Extensive tests with filters of 20 and 30 days widths have been carried out. These results show that the wider boxes show increasingly different $\tau_{\rm gran}$ values compared to the triangular smooth with a 10 day width. Also, results of $\tau_{\rm gran}$ using wider filters with a boxcar or a triangular shape are biased with respect to each other, while the results were consistent when using filters with 10 days widths. The 10 day filter width limits our sensitivity of $\tau_{\rm gran}$ to periods $<$~10 days, which is sufficiently long to study the granulation in red giants under the aforementioned assumption that these are of the order of hours up to a few days.

\section{Granulation parameters}
\label{sect:granparm}

Granulation parameters were first modeled in the Sun by \cite{1985ESASP.235..199H}. He approximated the autocovariance of the time evolution of the granulation by an exponential decay function with a characteristic time $\tau_{\rm gran}$. This results in a power spectrum with a Lorentzian profile:
\begin{equation}
P_{\rm H}(\nu)=\frac{4\sigma^2\tau_{\rm gran}}{1+(2\pi\nu\tau_{\rm gran})^\alpha}\ ,
\end{equation}
in which $P_{\rm H}(\nu)$ is the total power of the signal at frequency $\nu$, $\sigma$ is the characteristic amplitude of the granulation and $\alpha$ is a positive parameter characterizing the slope of the decay. We also define the amplitude of the granulation power, $P_{\rm gran}~=~4\sigma^2 \tau_{\rm gran}$. This approach has been successfully applied to the solar measurements by several instruments in both velocity and intensity \citep[e.g.][]{2008A&A...490.1143L}.

Other signals such as mesogranulation, supergranulation and active regions can be modeled with similar functions, and act at much longer time scales. We note here that it is mentioned in \cite{1985ESASP.235..199H} that the declining slopes of the functions may well be different from two. Different modified Harvey functions have indeed been used by for instance \cite{2004A&A...414.1139A}, \cite{2010MNRAS.402.2049H}, \cite{2009CoAst.160...74H}, \cite{2010A&A...522A...1K}, and in several methods used in this work, which are explained below in more detail. The use of values different from $\alpha=2$ were established empirically, but the physical meaning of the function with $\alpha$=2 has not been deeply investigated previously. 

Recalling that the autocorrelation of a signal is nothing but the Fourier
transform of its power spectrum, we can easily investigate the influence
of $\alpha$ on the autocorrelation function (ACF) of the temporal signal
of the granulation component. We note that, for a given $\tau$, an
increasing value for $\alpha$ indicates an increase in the temporal
correlation.
For $\alpha=2$, the ACF is obviously an exponential function decaying over
a time scale $\tau$.  When $\alpha$ tends to zero, the power spectrum
becomes flatter and flatter, i.e. tends to white noise. The ACF does then
decrease faster and faster, and tends to a Dirac function, typical for the
ACF of white noise. With increasing $\alpha$, the ACF broadens and converges to the sinc
function for $\alpha\rightarrow\infty$.%When $\alpha$ increases, the ACF becomes broader and
%tends to a sinc function.

To be able to compare characteristic time scales $\tau$ obtained with
different $\alpha$ values or even with different expressions of the
granulation spectrum, we define an effective time scale $\tau_\mathrm{eff}$ as the e-folding time of the ACF.
For $\alpha=2$, we thus recover $\tau=\tau_\mathrm{eff}$. In the following, we will study $\tau_{\rm eff}$ instead of $\tau_{\rm gran}$ to correctly account for slopes that differ from two.

%An increasing value for $\alpha$ indicates an increase in the temporal correlation of the signal. This can be understood from the following: when an exponentially damped process occurs stochastically, i.e., the signal can be modelled as $S(t)=s\otimes \mathcal N$ where $s(t)=e^{-t/\tau}$ and $\mathcal N$ is a random function, the power spectrum will be a Lorentzian ($\alpha$~=~2) multiplied by a random noise following a $\chi^2_{2\rm dof}$ law. In a sense there is no ``memory'' in the occurrence of the process, and the temporal correlation of the signal only comes from the ``afterglow'' of each event. In case the temporal signal is de-correlated we have random noise and $\alpha$~=~0. In the other extreme when the temporal correlation of the signal increases then $\alpha \rightarrow \infty$.

%An increased temporal correlation of the signal makes it more difficult to interpret the exact values of $\alpha$, $\tau_{\rm gran}$ and $\sigma$ in terms of their physical meaning. In a statistical sense, these parameters are however correlated with stellar parameters and global oscillation parameters and these correlations seem to be meaningful (see Section~4).

\subsection{Methods}

Six teams have fitted the granulation background of the red giants and the values of the parameters have been compared to check the validity of the results. For all the methods explained below, except CAN \citep{2010A&A...522A...1K}, the fit is performed by not taking into account the frequency region in which the solar-like oscillations are present. CAN fits the p-mode bump together with the background. The methods are all summarized in Tables~\ref{tbl-1} and \ref{tbl-1_1}.

%\noindent{\bf To sum up}

The OCT method fits the background of the power density spectrum (PDS) using one Harvey-like model:

\begin{equation}
P(\nu)=W+P_{\rm H}(\nu).
\label{eq2}
\end{equation}

\noindent The initial input parameters are [$P_{\rm gran,0}$, $\tau_{\rm gran,0}$, $\alpha_0$,  $W_0$] = [$P$, 0.01, 2, noise], in which $P$ is the maximum power of the binned power spectrum (bins vary with frequency and are 2$\Delta \nu$ wide), and noise is the minimum power of the binned power spectrum. All four parameters are left free in the non-linear least-squares fit. When 100 iterations are done, the model with the lowest $\chi^2$ is chosen. The standard deviations of the fitted parameters are used as the uncertainties. More details can be found in \cite{2010MNRAS.402.2049H}.
%When this fit does not converge we adapt one of the parameters (chosen randomly) with a percentage derived from a normally-distributed, floating-point, pseudo-random number with a mean of zero and a standard deviation of 0.3. When after 100 iterations with randomly adapted input parameters the fit still does not converge, we repeat the same procedure as described above but now with the noise fixed.

The method adopted by SYD \citep[see][for details]{2009CoAst.160...74H} consists of fitting the background with two modified Harvey functions \citep{karoffphd}:

\begin{equation}
P(\nu) = W + \sum_{i=1}^2 \frac{ 4\sigma_i^2 \tau_i} {1 + (2\pi \nu \tau_i)^2 + (2\pi \nu \tau_i)^4}\ ,
\end{equation}

\noindent with $\sigma_i$ the root mean square (rms) intensity in 
power density. $\tau_1$ corresponds to the granulation time scale and $\tau_2$ might correspond to another feature with no physical meaning so far.%To insure the correct fitting of the granulation component, the fitting had to be restricted to certain parts of the power spectrum. 
%The chosen ranges were as follows:
%\begin{enumerate}
%\item $\nu_{\rm max}  \rangle$ 25 $\mu$Hz: 1-280 muHz;
%\item $\nu_{\rm max}  \rangle$ 2 $\mu$Hz and $\nu_{\rm max}  \langle$ 25 $\mu$Hz: 1-100 muHz;
%\item$\nu_{\rm max}  \langle$ 2 muHz: 0.1-50 muHz (these are most likely not very reliable since the high-pass filter 
%affects all signal $\langle$ 1 muHz).
%\end{enumerate}
This method uses a Levenberg-Markward least-squares algorithm to fit the background and to estimate $\sigma_i$ and $\tau_i$. The initial guess value for the granulation time scale is scaled from the Sun using the relation $\tau_{\rm gran} \propto 1/\nu_{\rm max}$, where $\nu_{\rm max}$ is the frequency of maximum oscillation power and is determined before the background model is fitted. To stabilize the fit, only frequencies in the power spectrum lower than 2$\nu_{\rm max}$ are used for fitting the granulation component. This fit is performed on a binned PDS, leading to a Gaussian distribution of the points in the power spectrum. Monte-Carlo simulations are used to estimate uncertainties on the fitted parameters. For each simulation, a synthetic spectrum is generated by drawing random values following a $\chi^2$ distribution with two degree of freedom (d.o.f.) and repeating the fitting procedure described above. The uncertainties are taken as the standard deviations of the distributions derived from typically 1000 simulations.

%The CAN method fits the background with three Harvey laws and the fit of the $p$-mode bump with a Gaussian function as described in \cite{2010A&A...522A...1K}:

The CAN method uses three Harvey-like functions to model the stellar background signal and a Gaussian to account for the additional power due to p-mode oscillations:

\begin{equation}
P(\nu) = W + \sum_{i=1}^3 \frac{4\sigma_i^2\tau_i}{1+(2\pi\nu\tau_i)^4}+ P_{\rm g} {\rm exp}\Big (\frac {-(\nu_{\rm max} - \nu)^2}{2\sigma_g^2} \Big)\ ,
\end{equation}

\noindent with the condition $\tau_1 < \tau_2< \tau_3$, where the indices indicate consecutive background components. The Gaussian p-mode component has amplitude $P_{\rm g}$, central
frequency $\nu_{\rm max}$ and width $\sigma_{\rm g}$. This method uses a Bayesian Markov-Chain Monte-Carlo (MCMC) algorithm. 

%During the fitting process we kept $\nu_{\rm max}$ within $\pm$25\% of the value inferred from the visual inspection of the spectrum. The width of the power excess was allowed to vary between 5\% and 50\% of the initial guess of ?max.
\noindent The parameter $\tau_2$ corresponds to the granulation time scale. For now, a physical explanation of the other time scales, $\tau_1$ and $\tau_3$, is lacking. The latter could be related to a different scale of granulation and is currently under investigation. The granulation frequency, 1/$\tau_{\rm gran}=1/\tau_2$, was allowed to vary from 0 to 10 times $\nu_{\rm max}$. $P_{\rm g}$ was allowed to vary from zero to 10 times the average power in the spectrum around the initial guess for $\nu_{\rm max}$, and $W$ was kept between 0.5 and 2 times the average power at the high frequency end of the spectrum. The adopted likelihood function is based on a $\chi^2$ distribution with 2 d.o.f. and the uncertainties are determined from the marginalized posterior probability density distributions.

The DLB method (D. Buzasi, private communication) fits a single component Harvey model plus white noise as indicated in Eq.~\ref{eq2} with $\alpha=2$ as a fixed parameter.
This method uses a linear least squares fit over the frequency range [2/$T$, $\nu_{\rm Nyquist}$], wherer $T$ is the length of the time series. Initial guesses for the fit [$W$, $P_{\rm gran}$, $\tau_{\rm gran}$] are given by [$P_{\rm min}$, $P_{\rm max}$, 3~$\times 10^4$], where $P_{\rm min}$ and $P_{\rm max}$ are respectively the minimum and maximum power found in the region considered in the PDS. No uncertainties have been estimated with this method.

The COR method (B. Mosser, private communication) focusses on the frequency range around the break of the Harvey functions, just below the identified oscillation power. The fit of the background is performed with a single Harvey-like function in the frequency range between $\nu_{\rm max} / 30$ and  $ 2 \nu_{\rm max}$. The initial guesses of the fit are derived from the mean values of the background at $\nu_{\rm max}$ \citep{2011A&A...525L...9M,mosser2011} and at low frequency \citep[in the plateau region,][]{2010A&A...517A..22M}. The exponent of the Harvey component (Eq. 1) is a free parameter.%The exponent of the Harvey component (Eq. 1) can be fixed to 2 or let free. In this latter case, the mean value of the exponent is 2.10, with a 1-$\sigma$ dispersion limited to 0.14. Independent of the method, the variation of the ratio $P_{\rm gran} / P (\nu_{\rm max})$ with $\nu_{\rm max}$ is very limited. Its mean value is of about 23.}

With the A2Z method, the background is fitted with one Harvey-like model and a power law as described in \cite{2010A&A...511A..46M}:

\begin{equation}\label{eq:bgmodela2z}
 P(\nu)=W+P_{\rm H}(\nu)+a\nu^{-b}.
\end{equation}

\noindent The initial guess for $W$ is derived from a general relation between the stellar magnitude obtained from the KIC and the noise in the time series. This relation was determined from a larger sample of solar-like stars observed in short cadence by {\it Kepler}. The initial guess of the exponent $\alpha$ is 2, while that of $\tau_{\rm gran}$ is set to 2.1~$\times 10^{4}$~s. Assuming a $\chi^2$ distribution with 2 d.o.f., the fit is performed using a maximum likelihood estimator. The uncertainties are estimated by inverting the hessian matrix. 

For each method, we have computed $\tau_{\rm eff}$ converted from $\tau_{\rm gran}$ and $P_{\rm gran}$ as defined in Eq.~(1).

\subsection{Comparison of the results}

\begin{figure}[h!]
\epsscale{.80}
\includegraphics[angle=90,width=9cm]{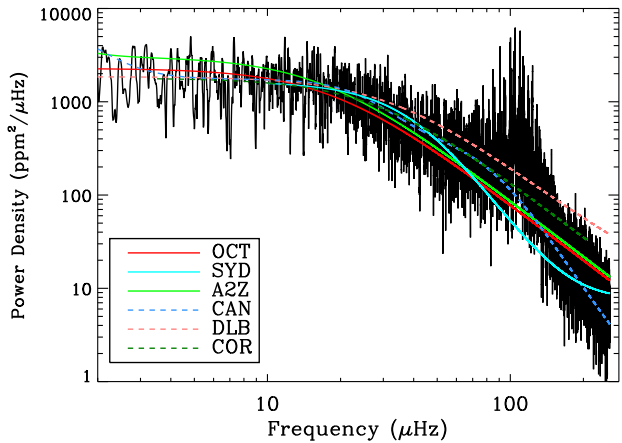}
\caption{Power density spectrum of a typical red giant, KIC~11618103, smoothed over 10 bins and with the background fitting results from all the different methods. \label{fig0}}
\end{figure}

Figure~\ref{fig0} shows a typical PDS of a red giant observed by {\it Kepler}. The results of each method listed above are shown and all of them provide the same qualitative results around the break of the granulation profile with small differences at very low frequency or at very high frequency. We note that one method can work well for one star while another method might work better for another star.

\begin{figure}[h!]
\epsscale{.80}
\includegraphics[angle=90,width=9cm]{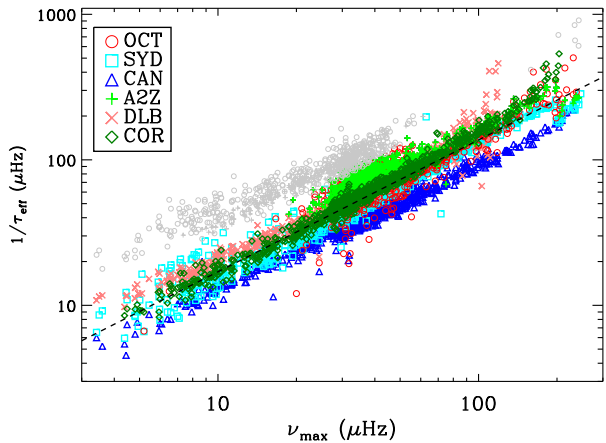}
\caption{Comparison of the relation between $\tau_{\rm eff}$ and $\nu_{\rm max}$ for the different methods. The dashed line represents the fit of  $\tau_{\rm eff}$ using all the methods together. The grey symbols represent the second branches of OCT and A2Z. \label{fig1}}
\end{figure}

\begin{figure}[htb]
%\epsscale{.80}
%\includegraphics[angle=90,width=8cm]{Figures/Agran_numax_Q0123.eps}
\includegraphics[angle=90,width=9cm]{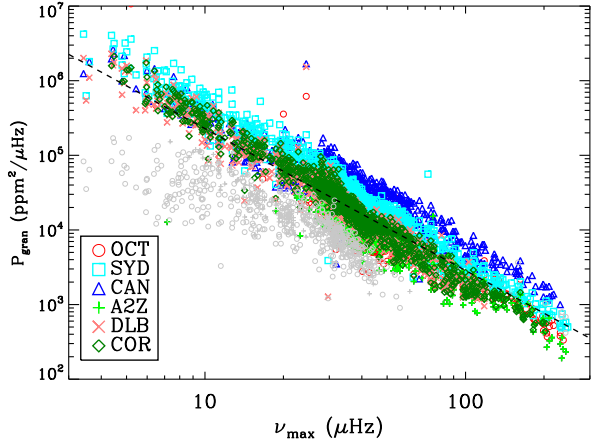}

\caption{Granulation power obtained with the background fitting, $P_{\rm gran}$ as a function of $\nu_{\rm max}$. Same legend as in Fig.~\ref{fig1}. The dashed line represents the fit of  $\tau_{\rm eff}$ using all the methods together.\label{fig2}}
\end{figure}

\begin{figure}[h!]
%\epsscale{.80}
%\includegraphics[angle=90,width=8cm]{Figures/comp_Agran_tau_gran_q012345_list1.eps}
\includegraphics[angle=90,width=9cm]{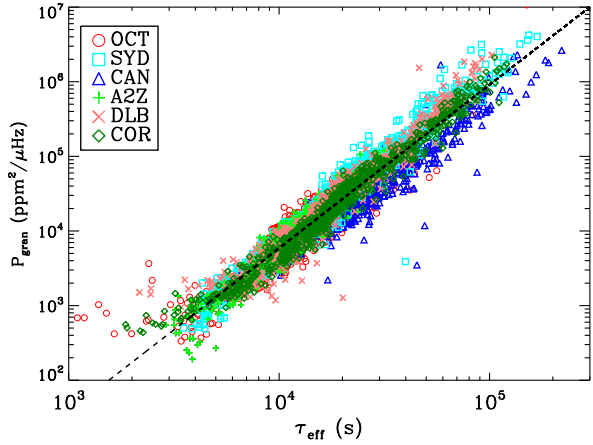}

\caption{Granulation power $P_{\rm gran}$ as a function of the granulation time scale, $\tau_{\rm eff}$. The dashed line indicates a linear fit through the results. The second branch of OCT and A2Z follows the same relation. \label{fig4}}
\end{figure}

Figures~\ref{fig1} and \ref{fig2} show the results of $\tau_{\rm eff}$ and $P_{\rm gran}$ for the methods described in Section 3.1. For the methods where more than one Harvey-like function was fitted, the one with the smallest time scale has been interpreted as belonging to the granulation for the SYD method and the middle time scale was used for the CAN method. The trends in the results from different methods are consistent, although they are not always the same. Interestingly the correlation between $\tau_{\rm eff}$ and $P_{\rm gran}$ as shown in Figure~\ref{fig4} is more coherent for the different methods. This could indicate a degeneracy in the fitted parameters coming from the use of different methods with different free parameters, models, data, number of components as it can be seen in Figure~\ref{fig0}. However for a given method, the results and the trends are consistent. %For the different methods, the average uncertainties on 1/$\tau_{\rm eff}$ are $\sim$~3.3, $\sim$~8.6, and $\sim$~9.3~$\mu$Hz for A2Z, CAN, and SYD respectively. Those seem rather small and might be underestimated.

For the different methods, the average uncertainties on 1/$\tau_{\rm eff}$ are 3\%, 9\%, 9\%, 1\%, and 6\% for A2Z, CAN, SYD, OCT, and COR respectively. Some of these uncertainties (in particular for A2Z, OCT and COR) seem rather small and might be underestimated so we assume that the uncertainty should rather be around 10\%. If we look at the dispersion between the methods for a given $\nu_{\rm max}$, 1/$\tau_{\rm eff}$ varies by $\sim$~30\% around the average value. The dispersion between the stars can be related to the different metallicities of the stars.

One thing that is apparent is the existence of a ``jump''  in the $\tau_{\rm gran}$ results at $\nu_{\rm max}$~$\sim$40~$\mu$Hz for two of the methods using a single power law, an initial guess for $\tau_{\rm gran}$ independent of $\nu_{\rm max}$ and $\alpha$ as a free parameter. This second branch is most likely due to the fact that the fit converges to a second feature in the power spectrum. The absence of the second branch in the results of DLB (guess of $\tau_{\rm gran}$ independent of $\nu_{\rm max}$ but fixed $\alpha$) indicates that the convergence is at least partly due to the fact that $\alpha$ is a free parameter.  We also performed tests with different initial guesses for $\tau_{\rm gran}$ (again independent of $\nu_{\rm max}$) and that reduces the second branch but the fit does not converge for high values of $\nu_{\rm max}$. Additionally, to investigate further the number of components, we performed tests with one method, A2Z in this case, in which we fitted either one or two Harvey components with several initial parameters to investigate how these affect the results (see the appendix for more details). It appears that by using two Harvey-like functions, we obtain only one branch instead of two branches (see discussion in Section 6). For all of these reasons, we think that on the one hand, the minimization algorithm of the fit could go to a local minimum instead of a global minimum, which can lead to the second branch. On the other hand, this second branch could be related to the initial guesses, for which we converge to a background component with lower $\tau_{\rm gran}$, which is likely to be something different from granulation. From the fact that the second branch also shows a strong correlation with $\nu_{\rm max}$, we expect it to be of stellar origin and it could correspond for instance to a phenomenon similar to bright points or stellar spots whose appearance is modulated by the stellar rotation.%To properly investigate the origin of this second branch, ? would be needed.}

As the results of the methods are qualitatively similar, we use one method for the representations of our results in the remainder of this work.

%{\bf The second branch...} We have investigated how we could explain the presence of a second upper branch of the A2Z and OCT methods. {\bf Check the values of the Sun...} 

\section{Granulation and other stellar parameters}
\label{sect:scalings}

%We studied if there were any relation between the granulation time scale and other parameters of the star, asteroseismic ones or fundamental ones. 
%Figure 3 shows that there is a relation with the position of maximum power of the modes ($\nu_{\rm max}$), which was also seen in Figure 2.
%So far no relation found with M and Teff. 
%With R and log g: see Figure2 and 3.
%Say a few words on the amplitude of granulation?

%Plots similar to the ones showed during KASC workshop...
\subsection{Granulation scaling relations and observations}

From some basic physical assumptions, it is possible to predict the
behavior of granulation with respect to stellar parameters.  As argued by
\citet{2009CoAst.160...74H} and \citet{2011A&A...529L...8K}, the convection cells travel a vertical distance that is proportional
to the pressure scale height, $H_p$, at a speed that is approximately
proportional to the sound speed, $c_s$.  Since $H_p
\propto T_{\rm eff}/g$ and $c_s \propto \sqrt{T_{\rm eff}}$, the
characteristic time scale of the granulation can be expressed as $\tau_{\rm
gran} \propto H_p/c_s \propto \sqrt{T_{\rm eff}}/g \propto
\sqrt{T_{\rm eff}}R^2/M \propto L/(M T_{\rm eff}^{3.5}) \propto 1/\nu_{\rm
max}$, where $g$ is surface gravity, $T_{\rm eff}$ is effective
temperature, $R$ is stellar radius, and $M$ is stellar mass.  This result
indicates that we should expect longer characteristic granulation
time scales on larger stars, i.e. stars with lower surface gravity.  Support for the argument is given by the 3D simulations by
\citet{trampedach:T-tau} carried out with the \citet{2000SoPh..192...91S} code. These simulations indicate that red giants have granulation cells that are roughly 15~$H_p$ wide.

The proportinality of the horizontal size of a granule, $d$ to the pressure scale height, $H_p$, 
\citep[e.g.,][]{1975ApJ...195..137S,1984ApJ...282..574A} implies the following: $d  \propto H_p \propto T_{\rm eff}/g \propto T_{\rm eff} R^2/M  \propto \sqrt{T_{\rm eff}}/\nu_{\rm max}$. Thus, the linear size of the granules is inversely proportional to the gravity, and since $T_{\rm eff}$ and the mass of the star vary less than the radii of red giants, it indicates that the size of the granules should increase with the radius of the star. 

For the number of granules on the surface, $N$, we derive the following: $N \propto R^2/d^2 \propto M^2/(T_{\rm eff}^2 R^2) \propto Mg/T_{\rm eff}^2$. This indicates that the number of granulation cells decreases with increasing stellar radius or equivalently decreasing surface gravity. So, large stars have a relatively small number of large granulation cells on their surfaces compared to smaller stars with a relatively higher number of smaller cells. These results are in line with earlier work by e.g. \citet{1975ApJ...195..137S,1984ApJ...282..574A}. Evidence for large granulation features has been observed on red
 supergiants such as Betelgeuse using interferometry (for a review see \cite{2003RPPh...66..789M}, for recent observations see \cite{2009A&A...508..923H,2009A&A...503..183O,2011A&A...529A.163O} and for their interpretation with 3D simulations, see  \cite{2010A&A...515A..12C}; \cite{2010ASPC..425..140K}).  We can test this hypothesis here also for red giants. 

As already pointed out by the original Harvey models, the power of the granulation $P_{\rm gran}$ is proportional to $\sigma^2\tau_{\rm gran}$, $\sigma$ being the rms intensity fluctuation. The proportionality with $\tau_{\rm gran}$ also means that $P_{\rm gran} \propto d$, i.e. fewer larger cells result in higher granulation power, and thus each cell has a larger influence on the total luminosity fluctuations of the star, and hence a larger influence on its variability. This has some similarity with the oscillations, i.e. the amplitudes of oscillation modes also increase with lower $\nu_{\rm max}$ \citep{2010A&A...517A..22M,2010ApJ...723.1607H}.

In Figures~\ref{fig1} and \ref{fig2} we fitted a power law for each parameter and for each method. The values of the slopes are listed in Table~\ref{tbl-2}. By taking the results of all the methods together, we find that $\tau_{\rm eff} \propto \nu_{\rm max}^{-0.89}$ while $P_{\rm gran} \propto \nu_{\rm max}^{-1.90}$, which is close to the relation derived above ($\tau_{\rm gran} \propto \nu_{\rm max}^{-1}$) and with the relations obtained by \citet{2011A&A...529L...8K} ($P_{\rm gran} \propto \nu_{\rm max}^{-2}$). Finally, it also agrees with analytical models of convection where a coefficient $-1$ is found for the $\tau_{\rm gran}$-$\nu_{\rm max}$ relation (R. Samadi private communication).
Given that both p modes and granulation are driven by convection, this correlation is not surprising.
%Given that both p modes and granulation phenomena are related to the convection, respectively to the excitation of the modes and the flows of gas, it is not surprising to have this correlation.

Figure~\ref{fig4} shows the variation of $P_{\rm gran}$ with $\tau_{\rm eff}$. The fit of a power law (see Table~\ref{tbl-2}) gives a slope of 2.18 for all the methods together, which agrees with the scaling laws. 

\begin{figure}[h!]
%\epsscale{.80}
%\includegraphics[angle=90,width=8cm]{Figures/comp_Agran_tau_gran_q012345_list1.eps}
\includegraphics[angle=90,width=9cm]{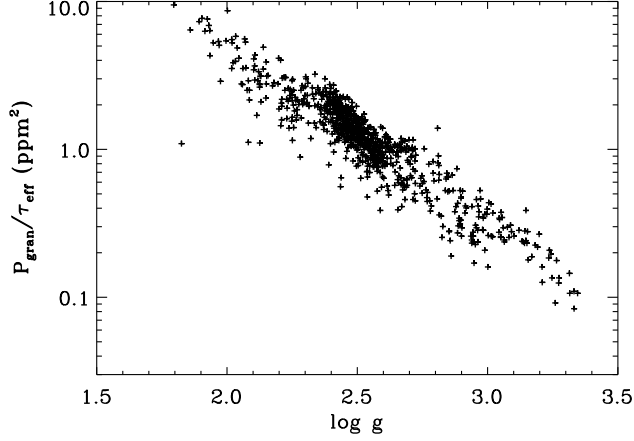}

\caption{Intensity variation obtained by the A2Z method as a function of log $g$ obtained by \citet{2010A&A...522A...1K}.\label{fig3}}
\end{figure}

It is interesting to study the variation of the ratio between  $P_{\rm gran}$ and $\tau_{\rm eff}$, which is proportional to the variance of the intensity variations. This quantity varies as a function of log $g$. This is illustrated in Figure~\ref{fig3}, where we used the granulation parameters obtained by A2Z as an example, but this tendency is qualitatively similar for the other methods. Stars that have smaller log $g$, thus larger radii present a higher intensity contrast compared to stars with a higher log $g$. 
This anti-correlation between $P_{\rm gran}/\tau_{\rm eff}$ and log $g$ seen in Figure~\ref{fig3} is mostly due to the fact that $P_{\rm gran}$ scales with $N^{-1}$ \citep{2011A&A...529L...8K},  which arises from the averaging of fluctuations over many (unresolved) granules, diminishing the power for large $N$, i.e, having fewer granules leads to less averaging over the stellar surface and the effect on luminosity from each granule is higher. %Note that for the two branches have disappeared here. It may suggest that $P_{\rm gran}$ compensates the fact that the code finds this jump.%Note that a small number of stars in our sample have large values of log $g$. As already mentioned in \cite{2010ApJ...723.1607H}, this can be due to the fact that more massive stars have a faster evolution, leading to a gap at this particular phase of the red giant branch (RGB).

%From the proportionality between we can also derive a value for the rms intensity variation to be: ???? {\bf It seems that $P_{\rm gran}/\tau_{\rm gran}$ is not a constant and that $\sigma^{2}$ varies with Teff and log g... But it is the opposite way compared to what I expected. See Figs.~\ref{fig3} and \ref{fig4}.}

\subsection{Granulation and global stellar parameters}

\begin{figure*}[htb]
%\epsscale{.80}
\begin{tabular}{cc}
\includegraphics[angle=90,width=9cm]{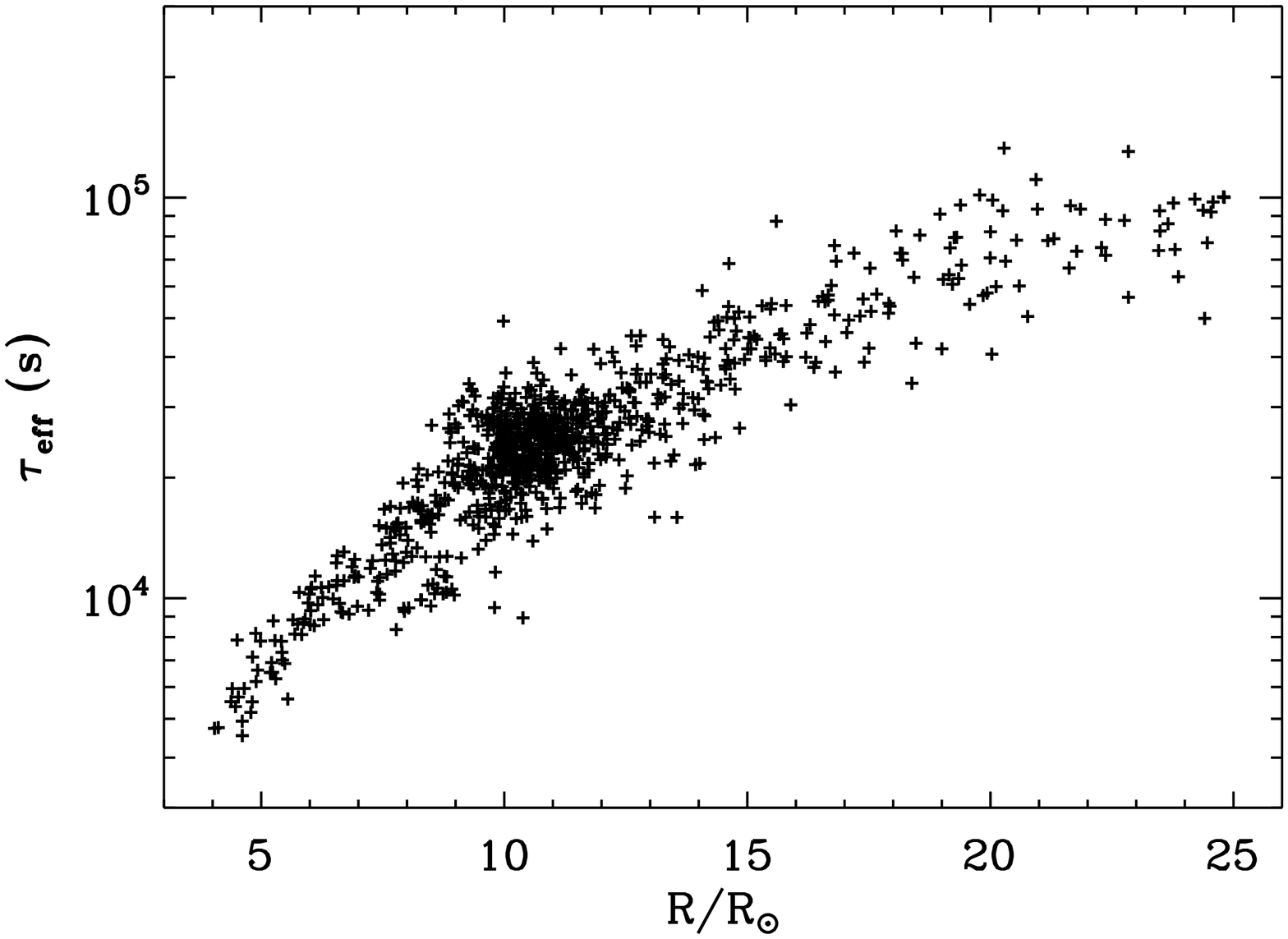}&\includegraphics[angle=90,width=9cm]{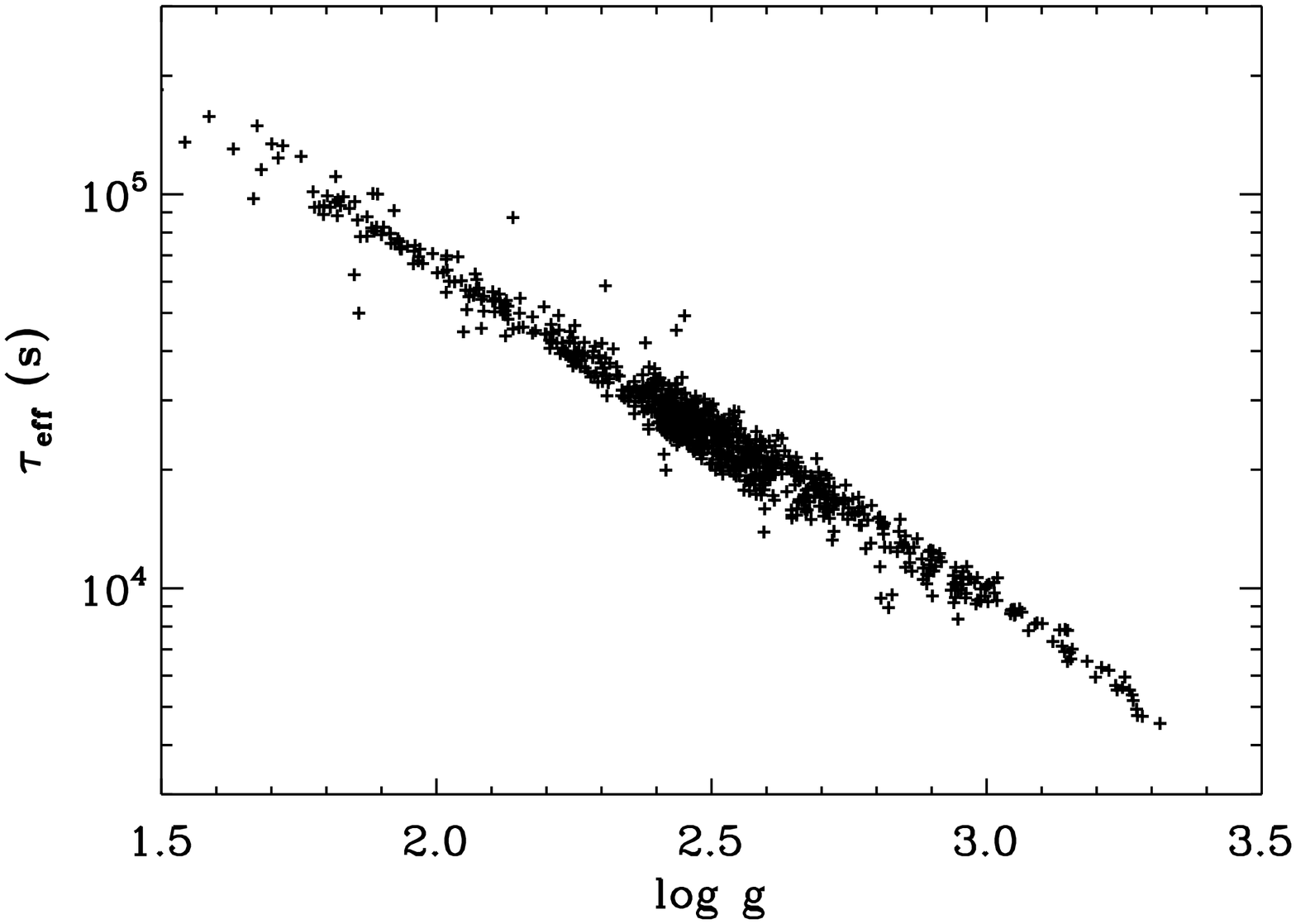}\\
\includegraphics[angle=90,width=9cm]{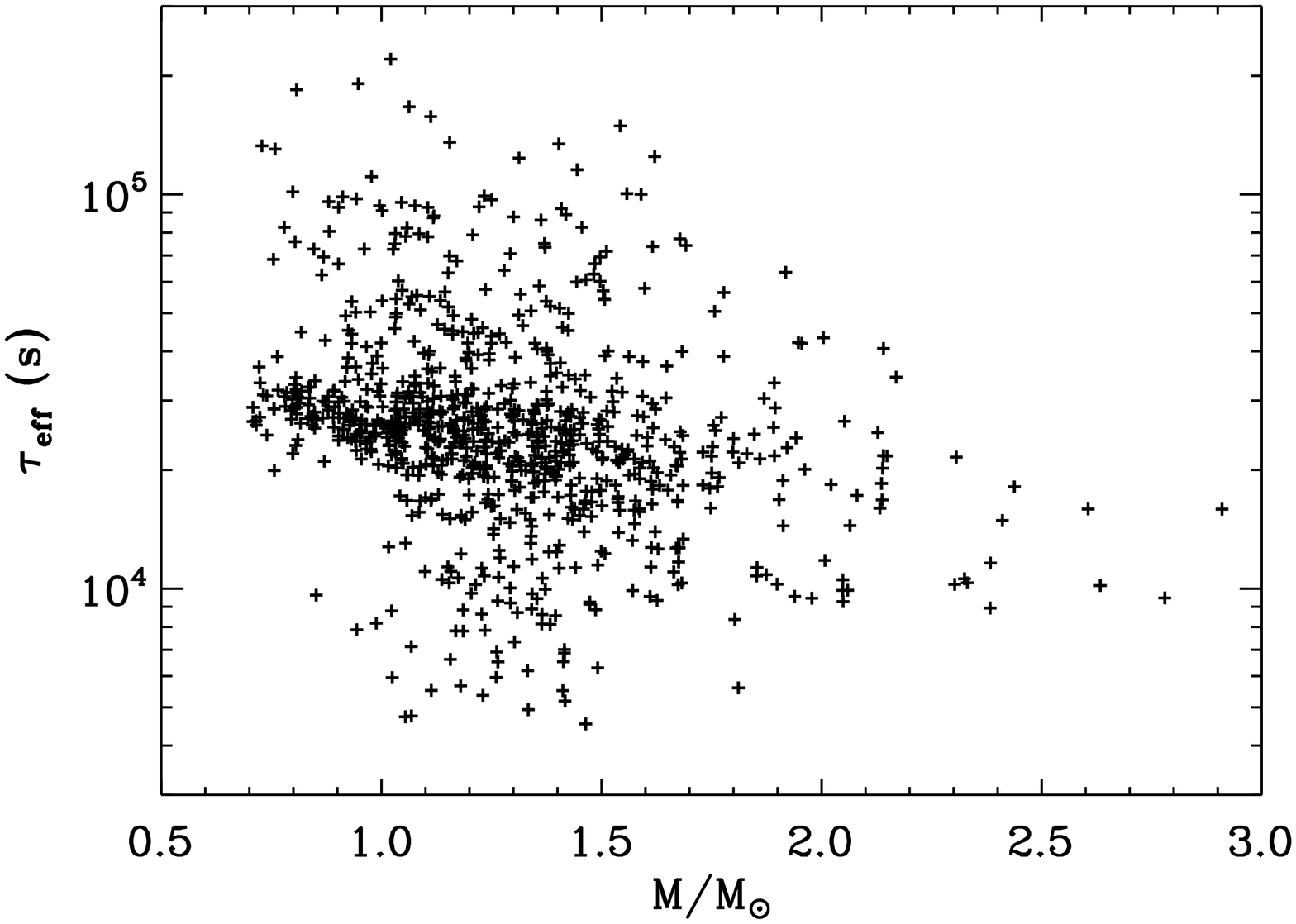}&\includegraphics[angle=90,width=9cm]{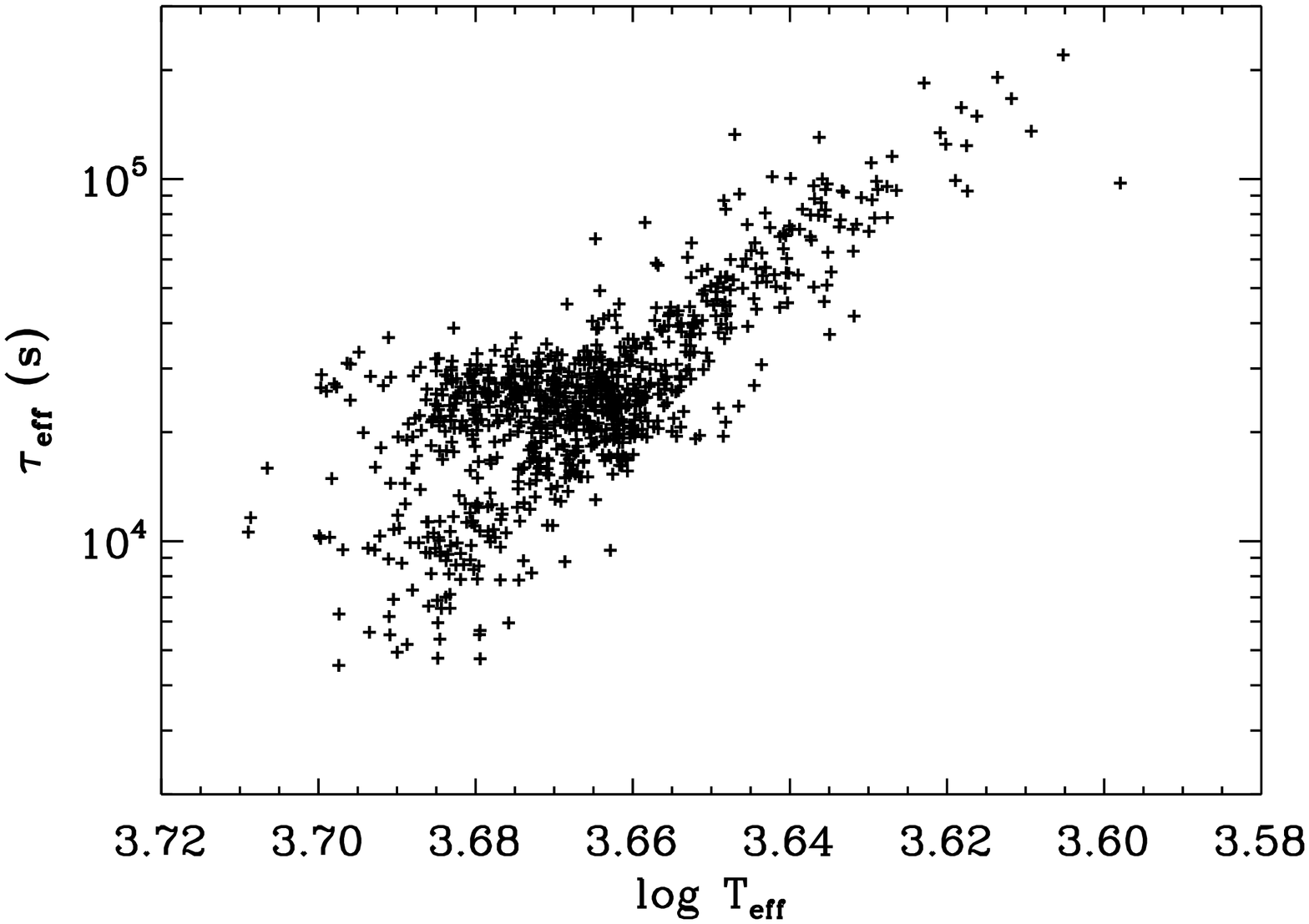}
\end{tabular}
\caption{Relation between the granulation time-scale, $\tau_{\rm eff}$, obtained by the CAN method and stellar parameters computed as described by \cite{2010A&A...522A...1K}.\label{fig5}}
\end{figure*}

\begin{figure*}[htb]
%\epsscale{.80}
\begin{tabular}{cc}
\includegraphics[angle=90,width=9cm]{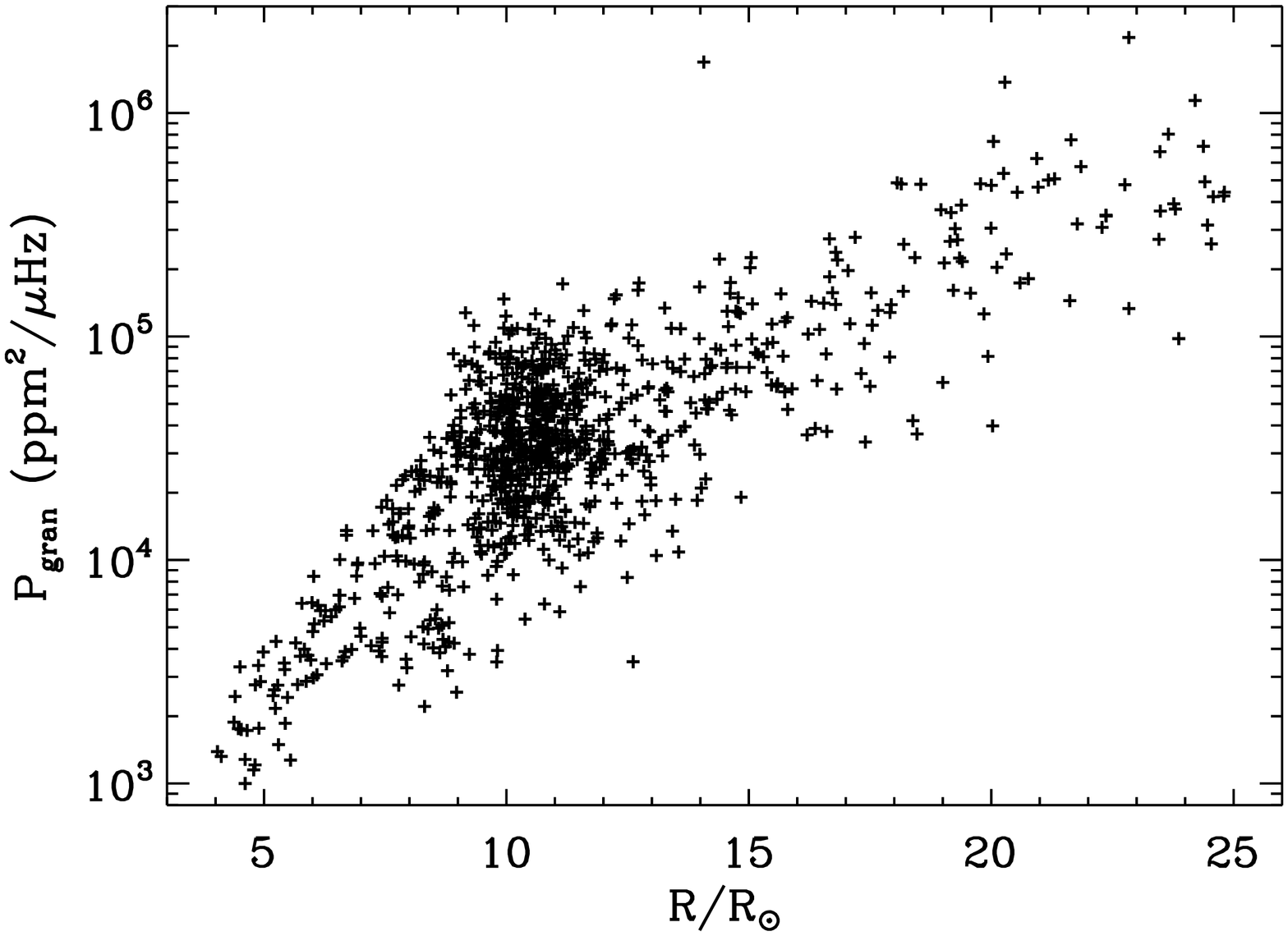}&\includegraphics[angle=90,width=9cm]{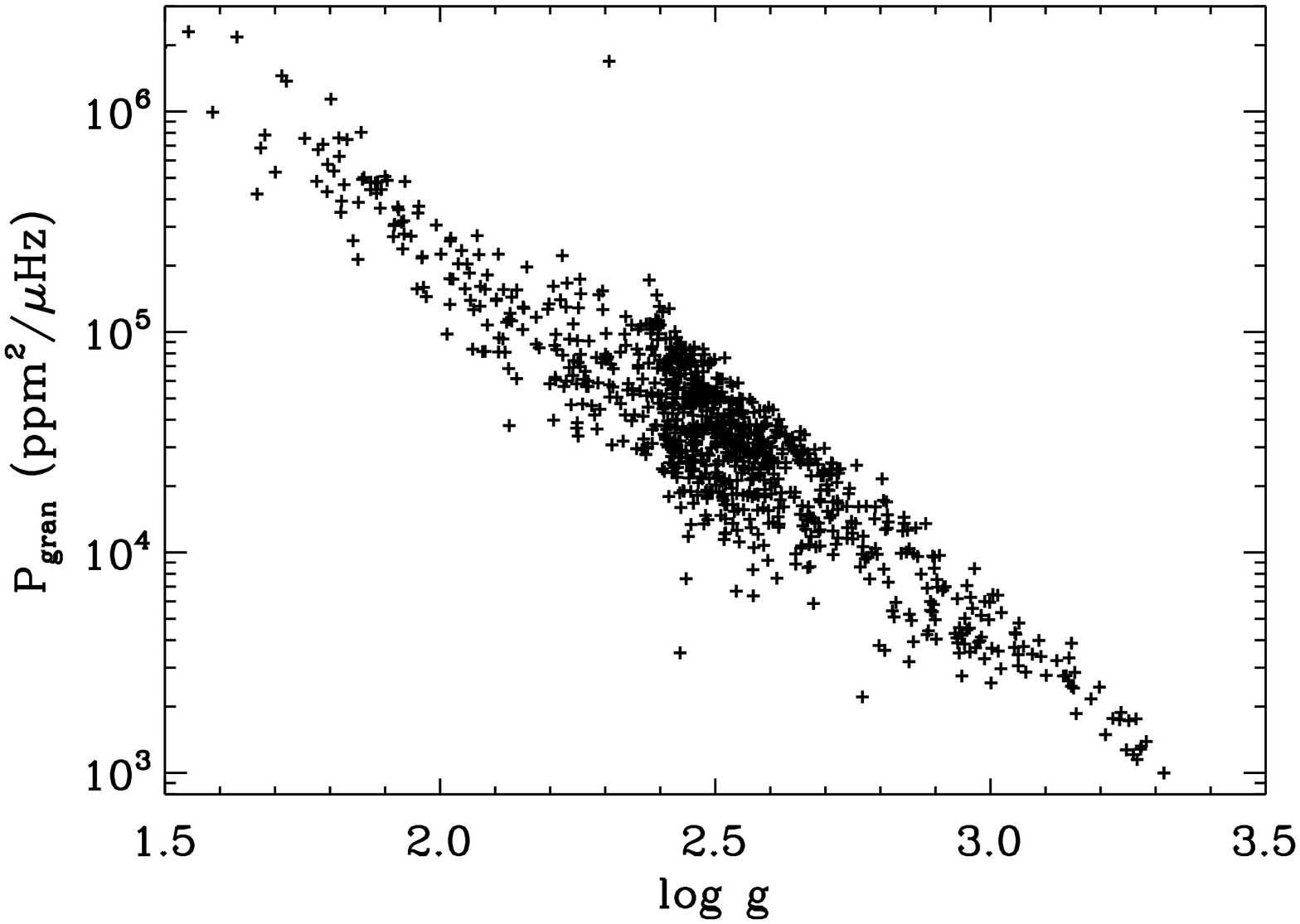}\\
\includegraphics[angle=90,width=9cm]{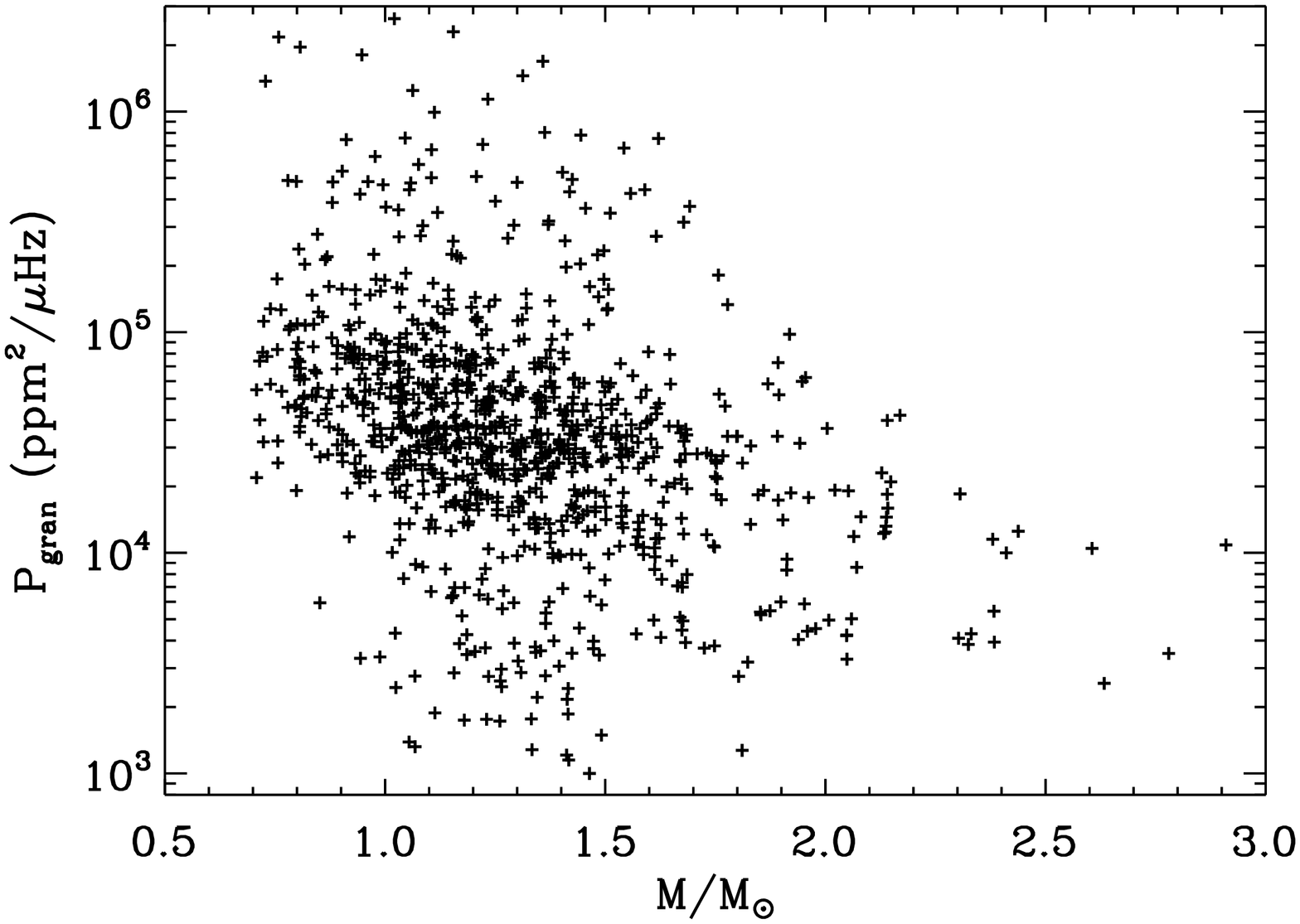}&\includegraphics[angle=90,width=9cm]{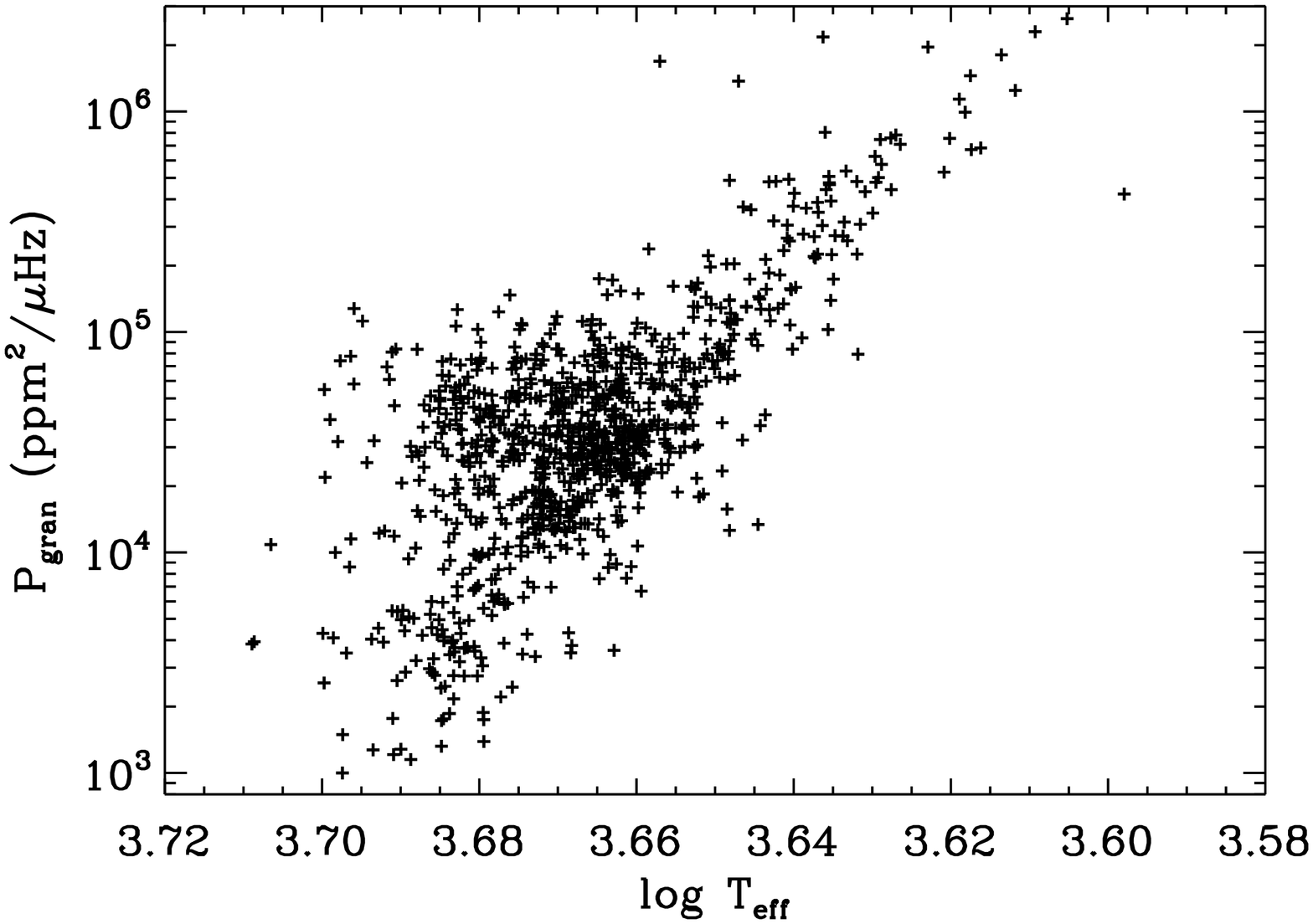}
\end{tabular}
\caption{Same as Figure~\ref{fig5} for the granulation power, $P_{\rm gran}$ obtained by the CAN method.\label{fig6}}
\end{figure*}

We investigate the dependence of the granulation parameters of the red-giant stars with stellar fundamental parameters R, M, $T_{\rm eff}$, and lo $g$. These parameters have been computed with the method described by \cite{2010A&A...522A...1K}  using 13 months long time series for a sample of 1035 red giants. Figure~\ref{fig5} shows how $\tau_{\rm eff}$ computed with method CAN varies with $R$, $\log g$, $M$, and $T_{\rm eff}$. The results from the CAN method are shown here because this method provides results of a large sample of stars. As predicted from scaling relations (see Section 4.1, $\tau_{\rm gran} \propto \sqrt{T_{\rm eff}}R^2/M$) for bigger stars, the granulation time scales are larger and we can fit a power law showing the correlation between $\tau_{\rm eff}$ and R as $R \propto\tau_{\rm eff}^{1.6}$. This is also consistent with the predicted anti-correlation between $\tau_{\rm gran}$ and $\log g$. Correlations of granulation parameters with mass and radius are not as tight as with $g=GM/R^2$, since the latter is the underlying independent variable \citep[see also][]{2011ApJ...730...63G}. The correlation with $T_{\rm eff}$ seems much less tight than with $\log g$. However, the plot of $\tau_{\rm eff}$ vs. $T_{\rm eff}$ is dominated by stellar evolution effects and hence shows a similar behavior as seen in the H-R diagram (Figure~\ref{fig8}) including the spread in points intrinsically linked to the distribution of stars in the H-R diagram. From the fact that $P_{\rm gran} \propto \tau_{\rm gran}$, we would expect similar correlations for the granulation power. These are indeed observed and shown in Figure~\ref{fig6}.

We note a bump around 10\,$R_{\odot}$, log $g$ = 2.5 and $T_{\rm eff}$ = 4700~K in Figures~\ref{fig5} and \ref{fig6}. It corresponds to the so-called red clump, corresponding to red giants that have gone through a Helium flash and are now in their He-core burning phase. In the plot showing $\tau_{\rm eff}$ as a function of $T_{\rm eff}$, we also clearly see the red-giant branch.

\subsection{Granulation in the H-R diagram}

Having a large number of red giants that are at different stages of their evolution allows us to study the distribution of the granulation parameters in this part of the H-R diagram.

\begin{figure*}[htb]
%\epsscale{.80}
\begin{tabular}{cc}
\includegraphics[angle=90,width=9cm]{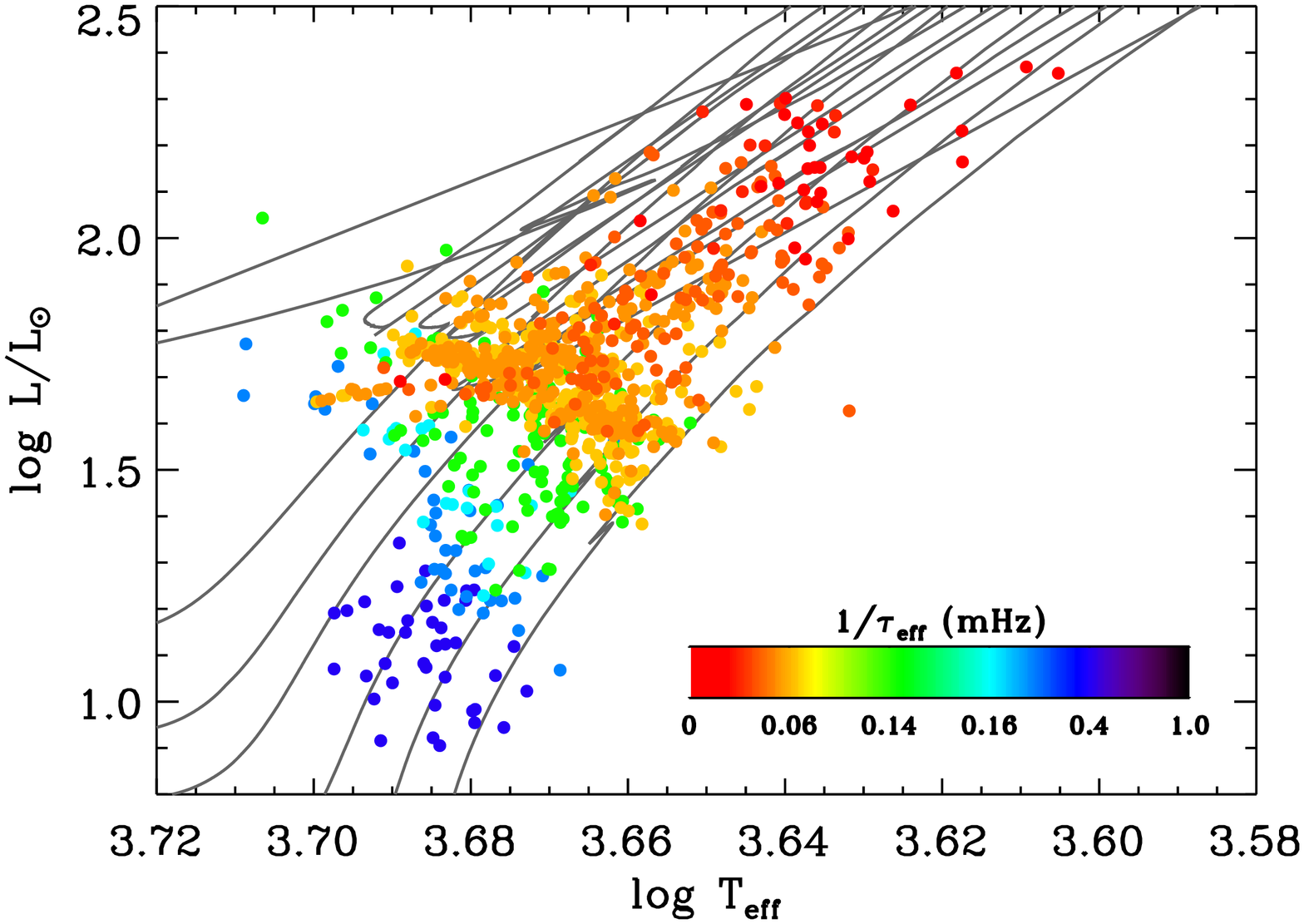}&\includegraphics[angle=90,width=9cm]{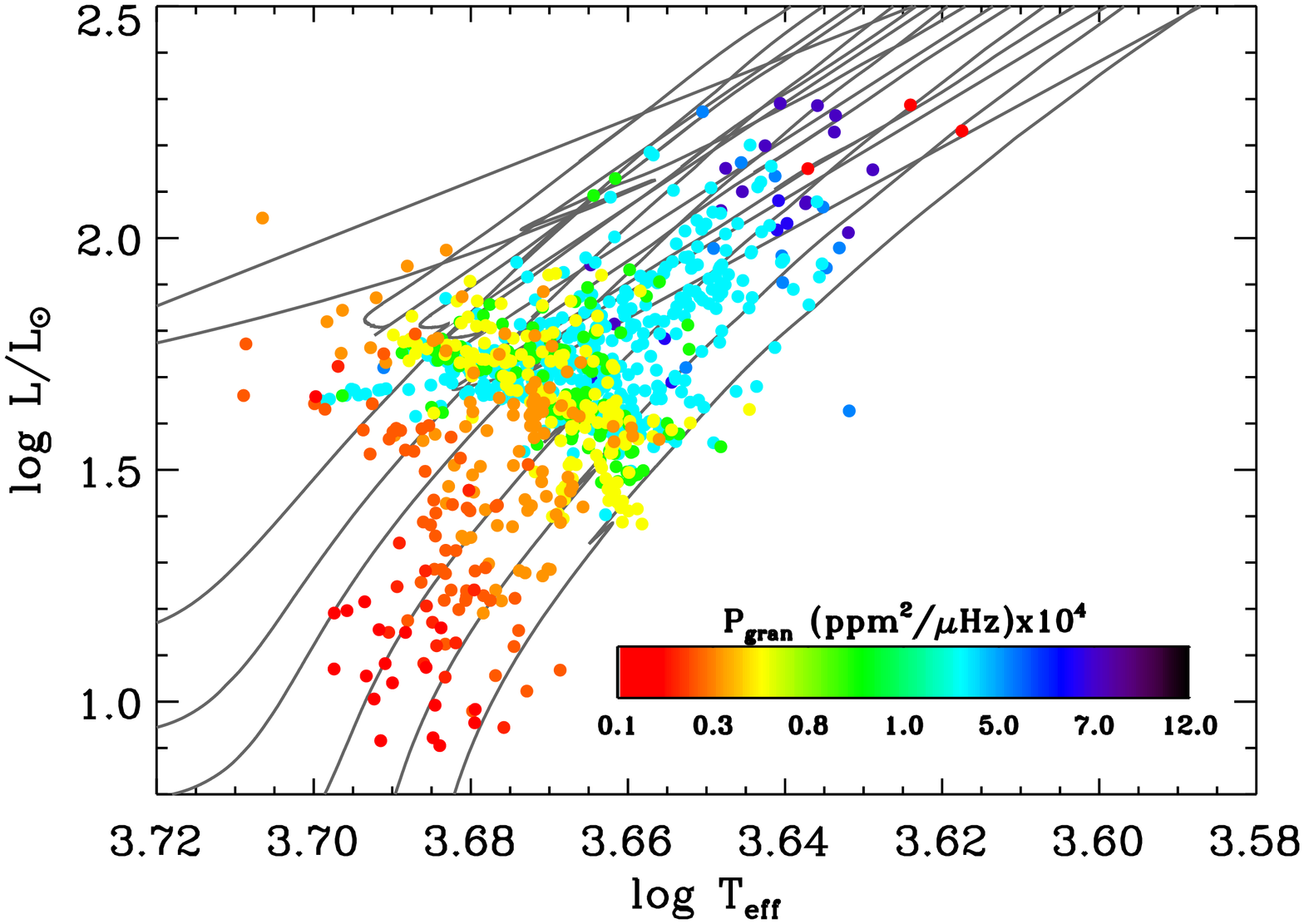}
\end{tabular}
\caption{For our sample of red giants, distribution of $1/\tau_{\rm eff}$ (left panel) and P$_{\rm gran}$ (right panel) shown with the color code in the H-R diagram. The grey lines represent BaSTI evolutionary tracks computed with solar metallicity. }
\label{fig8}
\end{figure*}

The left panel of Figure~\ref{fig8} shows the distribution of $1/\tau_{\rm eff}$ in our sample of red giants. As previously, the values of $T_{\rm eff}$ and $L$ have been computed as described by  \cite{2010A&A...522A...1K}. For stars with increasing luminosity, $1/\tau_{\rm eff}$ decreases, thus $\tau_{\rm eff}$ is larger, which also correlates with larger granulation cells. There is a large concentration of points at log\,$L/L_{\odot} \sim$~1.7. These stars belong to the red clump.%Interestingly, the granulation time scale of these stars $\tau_{\rm gran}$ is in the range 8000-20000~s, i.e. between 0.05 and 0.12~mHz . {\bf Is it coherent with the fact that these stars have similar radius and metallicity?}

The distribution of the granulation power, $P_{\rm gran}$, is shown in the right panel of Figure~\ref{fig8}. We note that low-luminosity stars have low $P_{\rm gran}$ values, which is once again similar to what we observe for the mode amplitudes \citep{2010A&A...517A..22M,mosser2011}. This is not surprising as the p modes are excited in the convection zone so the same order of magnitude of energy would be involved in both granulation and p-mode excitation. This is in agreement with the study of the height-to-background-ratio by \citet{mosser2011}. %For the granulation power as well the red clump is apparent. %, most of them corresponding to values between $5 \times 10^3$ and $4 \times 10^4$~$ppm^2 \mu Hz^{-1}$.

%In this analysis, we do not see any specific feature for the red bump or the secondary clump mentioned in \cite{2010ApJ...723.1607H} and \cite{2010A&A...522A...1K}. %Though there might be small hint in the granulation time-scale with values rather in the range 12500-2000~s. This needs to be confirmed.

Recently, \cite{2011Sci...332..205B}, \cite{2011Natur.471..608B}, and \cite{2011A&A...532A..86M} detected for the first time mixed modes in the red giants. They showed that the period spacings between the mixed modes make it possible to distinguish between stars ascending the red-giant branch, i.e., H-shell burning stars and stars in the red clump, i.e., He-core burning stars. Using this information, we investigated whether the stars show differences in granulation parameters related to their evolutionary phase. Due to the uncertainties in the granulation parameters a firm conclusion can not yet been drawn, although some indications of different granulation parameters with evolutionary state are present. It is at least clear that the red-clump stars have granulation values in a specific range, and the red-giant branch stars have their values spread over a larger range. It is not surprising as the structure of red-clump stars is very similar, which would lead to similar granulation characteristics. % It is more likely that $\tau_{\rm gran}$ and $P_{\rm gran}$ depend on the radius and the mass of the star rather than on its evolutionary stage.

%%%%%%%%%%%%%%%%%%%%%%%%%%%%%%%%%%%%%%%%%%%%%%%%%%%%%%%%%%%%%%%%%%%%%%%%%%%%%%%
%                                                                             %
%  Description, motivations and justifications for using the convection       %
%  simulations, as well as details about how we use them.                     %
%                                                                             %
%%%%%%%%%%%%%%%%%%%%%%%%%%%%%%%%%%%%%%%%%%%%%%%%%%%%%%%%%%%%%%%%%%%%%%%%%%%%%%%
%++++++++++++++++++++++++++++++++++++++++++++++++++++++++++++++++++++++++++++++
\section{3D Simulations of Convective Red Giant Atmospheres}
%++++++++++++++++++++++++++++++++++++++++++++++++++++++++++++++++++++++++++++++
\label{sect:rgsims}
To help us interpret the {\it Kepler} observations, we have turned to 3D hydrodynamic
simulations of convection in stellar surface layers. These simulations,
described in more detail in \citet{trampedach:T-tau}, were constructed for
direct comparison with observations and are therefore based on the state-of-the-art atomic physics. The thermodynamics is provided by tables of the
so-called MHD equation of state (EOS) \citep{mhd1,mhd2}, which is based on
explicit accounting for all excited states in all ions of the 16 most abundant
elements, and a
physical model of the (non-ideal) effect of interactions between particles.
The opacities are based on the Marcs stellar atmosphere
package \citep{b.gus}, but with several updates of the opacity sources and
line-opacities from the Atlas stellar atmosphere package
\citep{kur:line-data,kur:missolar}, as detailed in \citet{trampedach:T-tau}.
The three main processes governing the surface layers (i.e., what we can
observe) of late type stars are hydrodynamics, thermodynamics and radiative
transfer. These processes interact in very complicated ways, hence the need
for realistic first-principles simulations. The surface layers (by definition)
straddle the photospheric transition from the optically deep interior and the
optically thin atmosphere, which means the radiative transfer cannot be treated
in any known approximation. The full wavelength and ray-direction dependent
problem has to be considered and the simulations are based on the opacity
binning formulation introduced by \citet{aake:numsim1}.
The simulations are each performed on a grid of $150\times 150\times 82$
points, spanning about 13 pressure scale-heights in the vertical direction
(with 7 below the photosphere) and reaching up to an optical depth of
$\log_{10}\tau_{\rm Ross} = -4.5$. The horizontal extent primarily scales with
gravity, with only a slight increase with $T_{\rm eff}$ and is large enough to
cover about 10 major granules. Each simulation was relaxed to a statistically
steady state, with no systematic drifts in fluxes or mean structure over time. %{\bf How do you introduce the modes? How many?}
Surplus energy was extracted by artificially damping radial p modes during the
relaxation. After this relaxation, production runs were carried out, covering
at least 10 periods of the fundamental p mode. The resonant modes of the box
(the bottom is a node) are excited and damped by the convection in the box and
saturate at an amplitude given by the balance between the damping and driving.
With these fairly short time-series ranging from 14h55 for the coolest subgiant to 5d 17h30 for the hottest giant, we see two to three radial modes and similar
for the non-radial modes that has the box-width as horizontal wavelength. All 37 simulations of the grid
were performed for solar metallicity as in \citet{AG89} with He and Fe adjusted to mimick \citet{1993oee..conf...14G} and cover the zero-age main-sequence
from $T_{\rm eff}=4$\,300\,K to 6\,900\,K and up to giants of $\log g = 2.2$
between $T_{\rm eff}=5$\,000\,K and 4\,200\,K. Our sample of 1035 
{\it Kepler} red giants is spanned by seven of the 37 simulations of the grid.
About 11\% of the stars, however, fall outside the simulation grid and are
therefore not included in the comparisons of Section~\ref{sect:GranChar}.

Similar simulations that are not part of the grid, have been widely
  applied to solar and stellar observations. A study of solar Fe\,I
  and II lines by \citet{asplund:solar-Fe-shapes} showed remarkable
  agreement with the shape, asymmetry and wavelength-shift
  (without adjustable parameters to match these profiles to the
  observations). Note that 1D solar
  atmosphere models cannot reproduce these observations. A similar analysis was carried out for Procyon
  with a similar level of success.
  This agreement with detailed line-shapes, leads
  to a recognition of the Ni\,I blend with the [O]\,I line, leading
  to agreement between the oxygen abundances from [O]\,I and O\,I-lines
  \citep{prieto:SolarOI-abund}. The center-to-limb variation of Na-
  and O-lines in the Sun also matches observations \citep{prieto:sun-line-LD}.
  Even earlier simulations could reproduce the observed sizes, shapes
  and contrast of granules \citep{nordlundstein:RadDyn1991} and
  \citet{rosenthal:conv-osc} showed how the stratification change from a
  realistic 3D convective atmosphere accounts for most of the so-called
  \emph{surface term} in helioseismology.  Together,
  these disparate observational tests probe a large range of depths in the
  atmosphere of the simulation, giving us confidence that they are a better
  representation of stellar atmospheres than are 1D models. Hence,
  the 3D simulations have a larger predictive power.

%++++++++++++++++++++++++++++++++++++++++++++++++++++++++++++++++++++++++++++++
\subsection{``Observing'' the Simulations}
\begin{figure}[htb]

\includegraphics[height=6.5cm, trim=1cm 0 0 0]{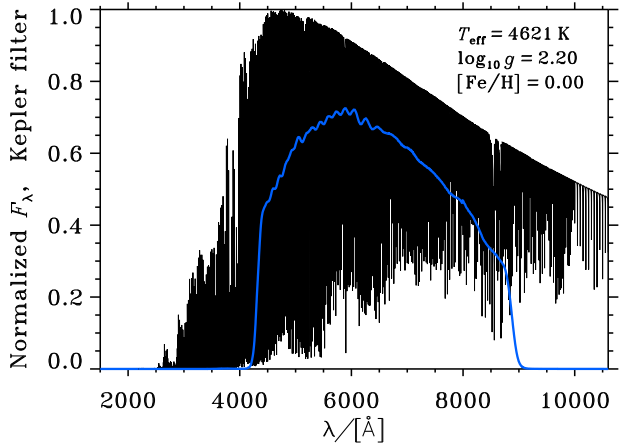}

\caption{Normalized emergent flux computed for a red giant simulation using the
opacity distribution functions as explained in Sect.~\ref{sect:simobs}.
The ``lines'' seen in this spectrum are therefore not actual spectral lines,
but 1100 ``giant'' lines, each quantifying the distribution of opacity in the
wavelength interval of each line (of about 20\,{\AA} in the optical).
The full calculation spans the spectrum from 860\,{\AA} to 20\,$\mu$m. The blue line represents the {\it Kepler} transmission filter.\label{figODF}}
\end{figure}

%++++++++++++++++++++++++++++++++++++++++++++++++++++++++++++++++++++++++++++++
\label{sect:simobs}
For each simulation, radiative transfer was performed for the full set of
wavelengths in the opacity distribution functions (ODFs). The ODFs that were
used are defined for 1100 wavelength regions, with 12 points in each of these
distribution functions.  We have artificially assigned wavelengths to these 12
points within their respective wavelength regions, in a way that gives them the
appropriate integration weight. The resulting spectra are saw-tooth shaped
with 20\,{\AA} wide bins in the optical. This is obviously not adequate for
monochromatic studies, but this wavelength re-ordering of opacity on a
20\,{\AA}-scale, has very little effect on broad-band colors.
Figure~\ref{figODF} shows such an ODF spectrum together with the {\it Kepler} filter. Having computed complete spectra for each snapshot of each simulation, and for
eight $\mu$-angles ($\mu=\cos\theta$ of the position angle on the stellar disk)
and four azimuthal $\phi$-angles, we then proceeded to average over
$\phi$ and convolve with the transmission curve of the {\it Kepler} filter. This results in {\it Kepler} intensities as function of $\mu$-angle and
time, $I(\mu,t)$.
%{\bfseriesadd reference: Bachtell \& Peters (2008)??? + Section 3.3 of InstrumentHandbook}

The subsequent transformation from specific intensities, to power spectra of
observable fluxes, was performed as described by \citet{trampedach:mons98} and later
described in great detail by \citet{hgl:conv-var}. Following the derivations
of \citet{hgl:conv-var} we write
\begin{equation}\label{eq:Fpower}
   P(\nu) = \frac{l^2}{2\pi R^2F^2}\sum_{ij} w_i\mu_i^2 w_j
                                               {\hat I}_{ij}{\hat I}_{ij}^*\ ,
\end{equation}

\noindent where $l$ is the horizontal extent of the simulation domain, ${\hat I}$ is the Fourier transform of $I$, ${\hat I}^*$ indicates the complex conjugate, $R$ is the radius
of the star and $2\pi R^2/l^2$ is the number of simulation domains that fit
over the visible half of the stellar surface. Division by the {\it Kepler} flux of the simulation, $F=\sum_i w_i\mu_i\sum_j I_{ij}$,
provides for the relative power spectra of the flux in Eq.~(\ref{eq:Fpower}).
The weights, $w_i$, and angles, $\mu_i$, of the angular quadrature are chosen
according to the method of \citet{radau:quadratur}, which gives an optimal set
of quadrature points, with the vertical direction included. The $\phi$-angles
that have index $j$, are equidistant and therefore have weights $1/N_\phi$.
${\hat I}_{ij}{\hat I}_{ij}^*$ is the
power of intensity in a particular direction, $\mu_i$ and $\phi_j$.

The power of the simulation
domain is diluted by the factor $2\pi R^2/l^2$ to account for the averaging out of 
uncorrelated convective fluctuations from different parts of the stellar disk.
The underlying assumption of this formulation is that patches on the stellar
surface, more than the distance $l$ apart are uncorrelated. The simulations
are dimensioned such that they each contain about a dozen major granules at any
one time, and the assumption is therefore reasonable.

The time series of the simulations are unfortunately not long enough to
sub-divide the time series in order to perform ensemble averaging to reduce
the noise. Instead we performed a running Gaussian smoothing of the power
spectra and these are then fitted with a Harvey-like model.
In the fitting process we also allow for a white noise component, and the two
strongest p modes. Due to the rather short time series, these modes are very
broad, and we fit them with Gaussian profiles. Only two or three modes
are visible in each simulation, and they are constrained to have a node at the
bottom of the simulation box, which means they don't correspond to real
eigenmodes of the star. Their frequencies, however, are still within the
p-mode bump of real stars.
%%%%%%%%%%%%%%%%%%%%%%%%%%%%%%%%%%%%%%%%%%%%%%%%%%%%%%%%%%%%%%%%%%%%%%%%%%%%%%%

%\begin{figure}[htb]

%\includegraphics[angle=90,width=9cm]{Figures/taueff_logg_simus_comp_obs_can.eps}
%\caption{Same as Figure~\ref{fig9} but as a function of log $g$.\label{fig10}}
%\end{figure}

\subsection{Results on the granulation characterization}
\label{sect:GranChar}
%Comparison with the sample of red giants observed. Comparison of the exponential law and the Harvey law fit?

%Before comparing the simulations with the {\it Kepler} observations, we did some tests to check that the exponential law used to fit the simulations agrees with the use of a Harvey-like model on a sample of 7 simulations. It came out that $\tau_{\rm gran}$ was similar within the uncertainties. The granulation power is not yet comparable between the observations and the simulations as we have a factor of four less power in the simulations, compared to the observations. This issue has yet to be resolved.

Having fitted the granulation power spectra of the simulations of the grid
to the generalized Harvey function of Eq.~(2), we now have
three granulation parameters as a function of $T_{\rm eff}$ and $\log g$. For
comparison with the red giants studied in this paper, we therefore interpolated
these parameters between the simulations, to the red giants based on their
atmospheric parameters, $T_{\rm eff}$ and $\log g$. Since the simulation grid
is irregular, we used linear interpolation on a  Delaunay triangulation of the
grid \citep{renka:triangulation}.

%We interpolated the granulation parameters of these 7 simulations to the red giants observed by {\it Kepler}, based on their $T_{\rm eff}$ and log $g$. {\bf Better explain that.}

\begin{figure}[htb]

\includegraphics[angle=90,height=6.5cm]{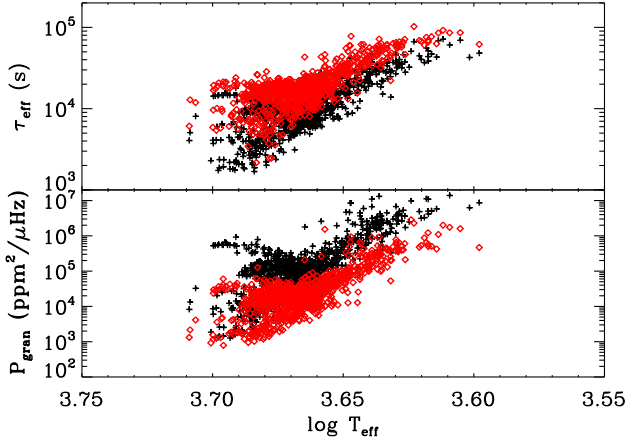}

\caption{Top panel: Granulation time scale, $\tau_{\rm eff}$, obtained with the 3D simulations vs. $T_{\rm eff}$ (black crosses) and interpolated to the observed sample of red giants. Results of the DLB method for the observations are over plotted (red diamonds). Bottom panel: Same as Top panel for the granulation power, $P_{\rm gran}$.\label{fig9}}
\end{figure}

\begin{figure}[htb]

\includegraphics[angle=90,width=9cm]{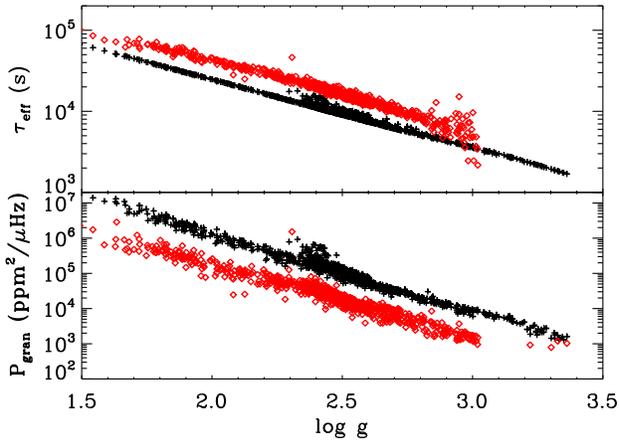}
\caption{Same as Figure~\ref{fig9} but as a function of log $g$.\label{fig10}}
\end{figure}

Top panels of Figures~\ref{fig9} and \ref{fig10} show how $\tau_{\rm eff}$ of the simulations varies with $T_{\rm eff}$ and log $g$. The simulations have the same trend as the observations. We observe very tight correlations between $\tau_{\rm eff}$ and log $g$. In the correlation between the granulation parameters  and $T_{\rm eff}$ (Figure~\ref{fig9})  the stellar evolution effects are again dominant (see also Section 4.2) and these figures show similar structure as seen in the H-R diagrams in Figure~\ref{fig8}. For $\tau_{\rm eff}$, the agreement between the observations and the simulations is better than a factor 2. We also compared $P_{\rm gran}$ from the simulations and the observations (bottom panels of Figures~\ref{fig9} and \ref{fig10}) and we note  a discrepancy of an order of magnitude. 

%\begin{figure}[htb]
%
%\includegraphics[angle=90,width=9cm]{Figures/Pgran_Teff_simus_comp_obs_can.eps}

%\caption{Same as Figure~\ref{fig9} for the granulation power, $P_{\rm gran}$. \label{fig11}}
%\end{figure}

%We note that the ``lower branch'' obtained with A2Z and OCT above 30~$\mu$Hz matches with the simulations.

%Finally, Figure~\ref{fig11} represents the distribution of the granulation time scale in both the simulations and the observations. The histograms are quite similar, specially if we take into account that the typical uncertainty associated to $\tau_{\rm gran}$ is of the order of 1$\times 10^3$~s, i.e. one bin.For this parameter, the simulations seem to reproduce well the observations. %{\bf TBC}

%\begin{figure}[htb]
%\includegraphics[angle=90,width=8cm]{Figures/histo_comp_tau_sim_obs_new.eps}
%\caption{Histogram of the values of $\tau_{\rm gran}$ from the observations (solid line) and from the simulations (dashed line).\label{fig11}}
%\end{figure}

%Info on physical processes.

\section{Discussion and conclusions}
\label{sect:conclusions}

%{\bf TBC}
%Talk about size of granules (observations + simulations), intensity contrast scaling with Agran...

For $\sim$~1000 red giants we studied 13 months of data obtained by the {\it Kepler} mission. These data were processed in order to reduce the instrumental effects while keeping as much information as possible to preserve the granulation signal.

{\bf Six} teams fitted the power spectra with different methods based on a Harvey model to estimate the granulation parameters. We first checked that the values obtained by the different methods were in general agreement. We notice that two branches appear for two of the methods that fitted only one Harvey model. In these cases, the methods are likely to fit for another feature present in the data. Judging from the correlation of the second branch with $\nu_{\rm max}$, we think this signal could also be of stellar origin. The comparison of the results of the red giants with the values we have for the Sun, suggests that the second branch would correspond to some high-frequency phenomena, such as bright points.
We regard a further investigation of this matter beyond the scope of this paper.

We found that the granulation time scale is proportional to $\nu_{\rm max}^{-0.89}$, while the granulation power is proportional to $\nu_{\rm max}^{-1.90}$, which agrees with theoretical scaling laws. From these results, we can use $\nu_{\rm max}$ as an initial guess for $\tau_{\rm gran}$ to fit the background.

We also showed that stars with larger radii have larger granules to the extent that their surfaces are covered by a smaller total number of granules.

Our study of the granulation power of red giants shows that larger stars present larger intensity fluctuations. Part of the reason for this is the smaller total number of granules covering the surface of larger stars, and hence less averaging out of fluctuations, compared to a star with many more (unresolved) granules. We also found that stars in the red clump have very similar values of granulation parameters.

Finally, we compared 3D simulations with our red giants sample and the granulation time scale is consistent with the observations in terms of trend with less than a factor of 2 difference.
The fact that for the observations and simulations, we have some dispersion in the plot $\tau_{\rm eff}$ vs. log $g$ can be due to several factors: 1) different stars having different metallicity (while our simulations have solar metallicity), 2) the instrumental noise that might have not been completely removed after the data processing, and 3) the radius is not necessarily determined with a sufficient accuracy for the interpolation of the simulations. The magnetic activity can also have an impact and could change the estimation of the granulation time scale by some factor, leading to different values of $\tau_{\rm eff}$ for stars that have similar $\nu_{\rm max}$.

%% The \notetoeditor{TEXT} command allows the author to communicate
%% information to the copy editor.  This information will appear as a
%% footnote on the printed copy for the manuscript style file.  Nothing will
%% appear on the printed copy if the preprint or
%% preprint2 style files are used.

%% The eqnarray environment produces multi-line display math. The end of
%% each line is marked with a \\. Lines will be numbered unless the \\
%% is preceded by a \nonumber command.
%% Alignment points are marked by ampersands (&). There should be two
%% ampersands (&) per line.


%% Included in this acknowledgments section are examples of the
%% AASTeX hypertext markup commands. Use \url without the optional [HREF]
%% argument when you want to print the url directly in the text. Otherwise,
%% use either \url or \anchor, with the HREF as the first argument and the
%% text to be printed in the second.

\acknowledgments
The authors gratefully acknowledge the \textit{Kepler} Science Team and all those who have contributed to making the \textit{Kepler} mission possible. Funding for the \textit{Kepler} Discovery mission is provided by NASAs Science Mission Directorate. NCAR is supported by the National Science Foundation. SH acknowledges financial support from the Netherlands Organisation for Scientific Research (NWO). This research was supported by grant AYA2010-17803 from the Spanish National Research Plan. RT was supported by NASA grant NNX08AI57G.

\bibliographystyle{apj}  
%\bibliography{apj-jour,/Users/Savita/Documents/BIBLIO_sav}
%\bibliography{add.bib}

\begin{thebibliography}{76}
\expandafter\ifx\csname natexlab\endcsname\relax\def\natexlab#1{#1}\fi

\bibitem[{{Aigrain} {et~al.}(2004){Aigrain}, {Favata}, \&
  {Gilmore}}]{2004A&A...414.1139A}
{Aigrain}, S., {Favata}, F., \& {Gilmore}, G. 2004, \aap, 414, 1139

\bibitem[{Anders \& Grevesse(1989)}]{AG89}
Anders, E., \& Grevesse, N. 1989, Geochim. Cosmochim. Acta, 53, 197

\bibitem[{{Antia} {et~al.}(1984){Antia}, {Chitre}, \&
  {Narasimha}}]{1984ApJ...282..574A}
{Antia}, H.~M., {Chitre}, S.~M., \& {Narasimha}, D. 1984, \apj, 282, 574

\bibitem[{{Asplund} {et~al.}(2000){Asplund}, {Ludwig}, {Nordlund}, \&
  {Stein}}]{2000A&A...359..669A}
{Asplund}, M., {Ludwig}, H.-G., {Nordlund}, {\AA}., \& {Stein}, R.~F. 2000,
  \aap, 359, 669

\bibitem[{Asplund {et~al.}(2000)Asplund, Nordlund, Trampedach, Prieto, \&
  Stein}]{asplund:solar-Fe-shapes}
Asplund, M., Nordlund, {\AA}., Trampedach, R., Prieto, C.~A., \& Stein, R.~F.
  2000, A\&A, 359, 729

\bibitem[{{Batalha} {et~al.}(2010){Batalha}, {Borucki}, {Koch}, {Bryson},
  {Haas}, {Brown}, {Caldwell}, {Hall}, {Gilliland}, {Latham}, {Meibom}, \&
  {Monet}}]{2010ApJ...713L.109B}
{Batalha}, N.~M., {et~al.} 2010, \apjl, 713, L109

\bibitem[{{Beck} {et~al.}(2011){Beck}, {Bedding}, {Mosser}, {Stello}, {Garcia},
  {Kallinger}, {Hekker}, {Elsworth}, {Frandsen}, {Carrier}, {De Ridder},
  {Aerts}, {White}, {Huber}, {Dupret}, {Montalb{\'a}n}, {Miglio}, {Noels},
  {Chaplin}, {Kjeldsen}, {Christensen-Dalsgaard}, {Gilliland}, {Brown},
  {Kawaler}, {Mathur}, \& {Jenkins}}]{2011Sci...332..205B}
{Beck}, P.~G., {et~al.} 2011, Science, 332, 205

\bibitem[{{Bedding} {et~al.}(2010){Bedding}, {Huber}, {Stello}, {Elsworth},
  {Hekker}, {Kallinger}, {Mathur}, {Mosser}, {Preston}, {Ballot}, {Barban},
  {Broomhall}, {Buzasi}, {Chaplin}, {Garc{\'{\i}}a}, {Gruberbauer}, {Hale}, {De
  Ridder}, {Frandsen}, {Borucki}, {Brown}, {Christensen-Dalsgaard},
  {Gilliland}, {Jenkins}, {Kjeldsen}, {Koch}, {Belkacem}, {Bildsten}, {Bruntt},
  {Campante}, {Deheuvels}, {Derekas}, {Dupret}, {Goupil}, {Hatzes}, {Houdek},
  {Ireland}, {Jiang}, {Karoff}, {Kiss}, {Lebreton}, {Miglio}, {Montalb{\'a}n},
  {Noels}, {Roxburgh}, {Sangaralingam}, {Stevens}, {Suran}, {Tarrant}, \&
  {Weiss}}]{2010ApJ...713L.176B}
{Bedding}, T.~R., {et~al.} 2010, \apjl, 713, L176

\bibitem[{{Bedding} {et~al.}(2011){Bedding}, {Mosser}, {Huber},
  {Montalb{\'a}n}, {Beck}, {Christensen-Dalsgaard}, {Elsworth},
  {Garc{\'{\i}}a}, {Miglio}, {Stello}, {White}, {De Ridder}, {Hekker}, {Aerts},
  {Barban}, {Belkacem}, {Broomhall}, {Brown}, {Buzasi}, {Carrier}, {Chaplin},
  {di Mauro}, {Dupret}, {Frandsen}, {Gilliland}, {Goupil}, {Jenkins},
  {Kallinger}, {Kawaler}, {Kjeldsen}, {Mathur}, {Noels}, {Aguirre}, \&
  {Ventura}}]{2011Natur.471..608B}
---. 2011, \nat, 471, 608

\bibitem[{{Bohm-Vitense}(1958)}]{bohm1958}
{Bohm-Vitense}, E. 1958, Z. Astrophys., 46, 108

\bibitem[{{Borucki} {et~al.}(2010){Borucki}, {Koch}, {Basri}, {Batalha},
  {Brown}, {Caldwell}, {Caldwell}, {Christensen-Dalsgaard}, {Cochran},
  {DeVore}, {Dunham}, {Dupree}, {Gautier}, {Geary}, {Gilliland}, {Gould},
  {Howell}, {Jenkins}, {Kondo}, {Latham}, {Marcy}, {Meibom}, {Kjeldsen},
  {Lissauer}, {Monet}, {Morrison}, {Sasselov}, {Tarter}, {Boss}, {Brownlee},
  {Owen}, {Buzasi}, {Charbonneau}, {Doyle}, {Fortney}, {Ford}, {Holman},
  {Seager}, {Steffen}, {Welsh}, {Rowe}, {Anderson}, {Buchhave}, {Ciardi},
  {Walkowicz}, {Sherry}, {Horch}, {Isaacson}, {Everett}, {Fischer}, {Torres},
  {Johnson}, {Endl}, {MacQueen}, {Bryson}, {Dotson}, {Haas}, {Kolodziejczak},
  {Van Cleve}, {Chandrasekaran}, {Twicken}, {Quintana}, {Clarke}, {Allen},
  {Li}, {Wu}, {Tenenbaum}, {Verner}, {Bruhweiler}, {Barnes}, \&
  {Prsa}}]{2010Sci...327..977B}
{Borucki}, W.~J., {et~al.} 2010, Science, 327, 977

\bibitem[{{Brown} {et~al.}(2011){Brown}, {Latham}, {Everett}, \&
  {Esquerdo}}]{2011arXiv1102.0342B}
{Brown}, T.~M., {Latham}, D.~W., {Everett}, M.~E., \& {Esquerdo}, G.~A. 2011,
  ArXiv e-prints 1102.0342

\bibitem[{{Canuto} \& {Mazzitelli}(1991)}]{1991ApJ...370..295C}
{Canuto}, V.~M., \& {Mazzitelli}, I. 1991, \apj, 370, 295

\bibitem[{{Canuto} \& {Mazzitelli}(1992)}]{1992ApJ...389..724C}
---. 1992, \apj, 389, 724

\bibitem[{{Chiavassa} {et~al.}(2010){Chiavassa}, {Haubois}, {Young}, {Plez},
  {Josselin}, {Perrin}, \& {Freytag}}]{2010A&A...515A..12C}
{Chiavassa}, A., {Haubois}, X., {Young}, J.~S., {Plez}, B., {Josselin}, E.,
  {Perrin}, G., \& {Freytag}, B. 2010, \aap, 515, A12

\bibitem[{{Ciardi} {et~al.}(2011){Ciardi}, {von Braun}, {Bryden}, {van Eyken},
  {Howell}, {Kane}, {Plavchan}, {Ram{\'{\i}}rez}, \&
  {Stauffer}}]{2011AJ....141..108C}
{Ciardi}, D.~R., {et~al.} 2011, \aj, 141, 108

\bibitem[{{Collet} {et~al.}(2007){Collet}, {Asplund}, \&
  {Trampedach}}]{2007A&A...469..687C}
{Collet}, R., {Asplund}, M., \& {Trampedach}, R. 2007, \aap, 469, 687

\bibitem[{{De Ridder} {et~al.}(2009){De Ridder}, {Barban}, {Baudin}, {Carrier},
  {Hatzes}, {Hekker}, {Kallinger}, {Weiss}, {Baglin}, {Auvergne}, {Samadi},
  {Barge}, \& {Deleuil}}]{2009Natur.459..398D}
{De Ridder}, J., {et~al.} 2009, \nat, 459, 398

\bibitem[{{Dupret} {et~al.}(2006){Dupret}, {Goupil}, {Samadi},
  {Grigahc{\`e}ne}, \& {Gabriel}}]{2006ESASP.624E..78D}
{Dupret}, M., {Goupil}, M., {Samadi}, R., {Grigahc{\`e}ne}, A., \& {Gabriel},
  M. 2006, in ESA Special Publication, Vol. 624, Proceedings of SOHO 18/GONG
  2006/HELAS I, Beyond the spherical Sun

\bibitem[{{Gai} {et~al.}(2011){Gai}, {Basu}, {Chaplin}, \&
  {Elsworth}}]{2011ApJ...730...63G}
{Gai}, N., {Basu}, S., {Chaplin}, W.~J., \& {Elsworth}, Y. 2011, \apj, 730, 63

\bibitem[{{Garc{\'{\i}}a} {et~al.}(2011){Garc{\'{\i}}a}, {Hekker}, {Stello},
  {Guti{\'e}rrez-Soto}, {Handberg}, {Huber}, {Karoff}, {Uytterhoeven},
  {Appourchaux}, {Chaplin}, {Elsworth}, {Mathur}, {Ballot},
  {Christensen-Dalsgaard}, {Gilliland}, {Houdek}, {Jenkins}, {Kjeldsen},
  {McCauliff}, {Metcalfe}, {Middour}, {Molenda-Zakowicz}, {Monteiro}, {Smith},
  \& {Thompson}}]{2011MNRAS.414L...6G}
{Garc{\'{\i}}a}, R.~A., {et~al.} 2011, \mnras, 414, L6

\bibitem[{{Girardi} \& {Salaris}(2001)}]{2001MNRAS.323..109G}
{Girardi}, L., \& {Salaris}, M. 2001, \mnras, 323, 109

\bibitem[{{Gough}(1977{\natexlab{a}})}]{1977LNP....71...15G}
{Gough}, D. 1977{\natexlab{a}}, in Lecture Notes in Physics, Berlin Springer
  Verlag, Vol.~71, Problems of Stellar Convection, ed. {E.~A.~Spiegel \&
  J.-P.~Zahn}, 15--56

\bibitem[{{Gough}(1977{\natexlab{b}})}]{1977ApJ...214..196G}
{Gough}, D.~O. 1977{\natexlab{b}}, \apj, 214, 196

\bibitem[{{Grevesse} \& {Noels}(1993)}]{1993oee..conf...14G}
{Grevesse}, N., \& {Noels}, A. 1993, in Origin and evolution of the elements:
  proceedings of a symposium in honour of H. Reeves, held in Paris, June 22-25,
  1992. Edited by N. Prantzos, E. Vangioni-Flam and M. Casse. Published by
  Cambridge University Press, Cambridge, England, 1993, p.14, ed.
  N.~{Prantzos}, E.~{Vangioni-Flam}, \& M.~{Casse}, 14

\bibitem[{Gustafsson(1973)}]{b.gus}
Gustafsson, B. 1973in  (Uppsala: Landstingets Verkst{\"a}der), 1--31

\bibitem[{{Harvey}(1985)}]{1985ESASP.235..199H}
{Harvey}, J. 1985, in ESA Special Publication, Vol. 235, Future Missions in
  Solar, Heliospheric \& Space Plasma Physics, ed. E.~{Rolfe} \& B.~{Battrick},
  199

\bibitem[{{Haubois} {et~al.}(2009){Haubois}, {Perrin}, {Lacour}, {Verhoelst},
  {Meimon}, {Mugnier}, {Thi{\'e}baut}, {Berger}, {Ridgway}, {Monnier},
  {Millan-Gabet}, \& {Traub}}]{2009A&A...508..923H}
{Haubois}, X., {et~al.} 2009, \aap, 508, 923

\bibitem[{{Hekker} {et~al.}(2009){Hekker}, {Kallinger}, {Baudin}, {De Ridder},
  {Barban}, {Carrier}, {Hatzes}, {Weiss}, \& {Baglin}}]{2009A&A...506..465H}
{Hekker}, S., {et~al.} 2009, \aap, 506, 465

\bibitem[{{Hekker} {et~al.}(2010){Hekker}, {Broomhall}, {Chaplin}, {Elsworth},
  {Fletcher}, {New}, {Arentoft}, {Quirion}, \&
  {Kjeldsen}}]{2010MNRAS.402.2049H}
---. 2010, \mnras, 402, 2049

\bibitem[{{Hekker} {et~al.}(2011{\natexlab{a}}){Hekker}, {Basu}, {Stello},
  {Kallinger}, {Grundahl}, {Mathur}, {Garc{\'{\i}}a}, {Mosser}, {Huber},
  {Bedding}, {Szab{\'o}}, {De Ridder}, {Chaplin}, {Elsworth}, {Hale},
  {Christensen-Dalsgaard}, {Gilliland}, {Still}, {McCauliff}, \&
  {Quintana}}]{2011A&A...530A.100H}
---. 2011{\natexlab{a}}, \aap, 530, A100

\bibitem[{{Hekker} {et~al.}(2011{\natexlab{b}}){Hekker}, {Elsworth}, {De
  Ridder}, {Mosser}, {Garc{\'{\i}}a}, {Kallinger}, {Mathur}, {Huber}, {Buzasi},
  {Preston}, {Hale}, {Ballot}, {Chaplin}, {R{\'e}gulo}, {Bedding}, {Stello},
  {Borucki}, {Koch}, {Jenkins}, {Allen}, {Gilliland}, {Kjeldsen}, \&
  {Christensen-Dalsgaard}}]{2011A&A...525A.131H}
---. 2011{\natexlab{b}}, \aap, 525, A131

\bibitem[{{Herschel}(1801)}]{1801RSPT...91..265H}
{Herschel}, W. 1801, Royal Society of London Philosophical Transactions Series
  I, 91, 265

\bibitem[{{Huber} {et~al.}(2009){Huber}, {Stello}, {Bedding}, {Chaplin},
  {Arentoft}, {Quirion}, \& {Kjeldsen}}]{2009CoAst.160...74H}
{Huber}, D., {Stello}, D., {Bedding}, T.~R., {Chaplin}, W.~J., {Arentoft}, T.,
  {Quirion}, P., \& {Kjeldsen}, H. 2009, Communications in Asteroseismology,
  160, 74

\bibitem[{{Huber} {et~al.}(2010){Huber}, {Bedding}, {Stello}, {Mosser},
  {Mathur}, {Kallinger}, {Hekker}, {Elsworth}, {Buzasi}, {De Ridder},
  {Gilliland}, {Kjeldsen}, {Chaplin}, {Garc{\'{\i}}a}, {Hale}, {Preston},
  {White}, {Borucki}, {Christensen-Dalsgaard}, {Clarke}, {Jenkins}, \&
  {Koch}}]{2010ApJ...723.1607H}
{Huber}, D., {et~al.} 2010, \apj, 723, 1607

\bibitem[{Hummer \& Mihalas(1988)}]{mhd1}
Hummer, D.~G., \& Mihalas, D. 1988, ApJ, 331, 794

\bibitem[{{Jenkins} {et~al.}(2010){Jenkins}, {Caldwell}, {Chandrasekaran},
  {Twicken}, {Bryson}, {Quintana}, {Clarke}, {Li}, {Allen}, {Tenenbaum}, {Wu},
  {Klaus}, {Middour}, {Cote}, {McCauliff}, {Girouard}, {Gunter}, {Wohler},
  {Sommers}, {Hall}, {Uddin}, {Wu}, {Bhavsar}, {Van Cleve}, {Pletcher},
  {Dotson}, {Haas}, {Gilliland}, {Koch}, \& {Borucki}}]{2010ApJ...713L..87J}
{Jenkins}, J.~M., {et~al.} 2010, \apjl, 713, L87

\bibitem[{{Kallinger} {et~al.}(2010{\natexlab{a}}){Kallinger}, {Mosser},
  {Hekker}, {Huber}, {Stello}, {Mathur}, {Basu}, {Bedding}, {Chaplin}, {De
  Ridder}, {Elsworth}, {Frandsen}, {Garc{\'{\i}}a}, {Gruberbauer}, {Matthews},
  {Borucki}, {Bruntt}, {Christensen-Dalsgaard}, {Gilliland}, {Kjeldsen}, \&
  {Koch}}]{2010A&A...522A...1K}
{Kallinger}, T., {et~al.} 2010{\natexlab{a}}, \aap, 522, A1

\bibitem[{{Kallinger} {et~al.}(2010{\natexlab{b}}){Kallinger}, {Weiss},
  {Barban}, {Baudin}, {Cameron}, {Carrier}, {De Ridder}, {Goupil},
  {Gruberbauer}, {Hatzes}, {Hekker}, {Samadi}, \&
  {Deleuil}}]{2010A&A...509A..77K}
---. 2010{\natexlab{b}}, \aap, 509, A77

\bibitem[{{Karoff}(2008)}]{karoffphd}
{Karoff}, C. 2008, PhD thesis

\bibitem[{{Kiss} {et~al.}(2010){Kiss}, {Monnier}, {Bedding}, {Tuthill}, {Zhao},
  {Ireland}, \& {Ten Brummelaar}}]{2010ASPC..425..140K}
{Kiss}, L.~L., {Monnier}, J.~D., {Bedding}, T.~R., {Tuthill}, P., {Zhao}, M.,
  {Ireland}, M.~J., \& {Ten Brummelaar}, T.~A. 2010, in Astronomical Society of
  the Pacific Conference Series, Vol. 425, Hot and Cool: Bridging Gaps in
  Massive Star Evolution, ed. {C.~Leitherer, P.~Bennett, P.~Morris, \& J.~van
  Loon}, 140

\bibitem[{{Kjeldsen} \& {Bedding}(2011)}]{2011A&A...529L...8K}
{Kjeldsen}, H., \& {Bedding}, T.~R. 2011, \aap, 529, L8

\bibitem[{{Koch} {et~al.}(2010){Koch}, {Borucki}, {Basri}, {Batalha}, {Brown},
  {Caldwell}, {Christensen-Dalsgaard}, {Cochran}, {DeVore}, {Dunham},
  {Gautier}, {Geary}, {Gilliland}, {Gould}, {Jenkins}, {Kondo}, {Latham},
  {Lissauer}, {Marcy}, {Monet}, {Sasselov}, {Boss}, {Brownlee}, {Caldwell},
  {Dupree}, {Howell}, {Kjeldsen}, {Meibom}, {Morrison}, {Owen}, {Reitsema},
  {Tarter}, {Bryson}, {Dotson}, {Gazis}, {Haas}, {Kolodziejczak}, {Rowe}, {Van
  Cleve}, {Allen}, {Chandrasekaran}, {Clarke}, {Li}, {Quintana}, {Tenenbaum},
  {Twicken}, \& {Wu}}]{2010ApJ...713L..79K}
{Koch}, D.~G., {et~al.} 2010, \apjl, 713, L79

\bibitem[{Kurucz(1992{\natexlab{a}})}]{kur:line-data}
Kurucz, R.~L. 1992{\natexlab{a}}, Rev. Mex. Astron. Astrofis., 23, 45

\bibitem[{Kurucz(1992{\natexlab{b}})}]{kur:missolar}
---. 1992{\natexlab{b}}, Rev. Mex. Astron. Astrofis., 23, 181

\bibitem[{{Lefebvre} {et~al.}(2008){Lefebvre}, {Garc{\'{\i}}a},
  {Jim{\'e}nez-Reyes}, {Turck-Chi{\`e}ze}, \& {Mathur}}]{2008A&A...490.1143L}
{Lefebvre}, S., {Garc{\'{\i}}a}, R.~A., {Jim{\'e}nez-Reyes}, S.~J.,
  {Turck-Chi{\`e}ze}, S., \& {Mathur}, S. 2008, \aap, 490, 1143

\bibitem[{{Ludwig} {et~al.}(2009){Ludwig}, {Samadi}, {Steffen}, {Appourchaux},
  {Baudin}, {Belkacem}, {Boumier}, {Goupil}, \& {Michel}}]{2009A&A...506..167L}
{Ludwig}, H., {et~al.} 2009, \aap, 506, 167

\bibitem[{Ludwig(2006)}]{hgl:conv-var}
Ludwig, H.-G. 2006, A\&A, 445, 661

\bibitem[{{Mathur} {et~al.}(2010){Mathur}, {Garc{\'{\i}}a}, {R{\'e}gulo},
  {Creevey}, {Ballot}, {Salabert}, {Arentoft}, {Quirion}, {Chaplin}, \&
  {Kjeldsen}}]{2010A&A...511A..46M}
{Mathur}, S., {et~al.} 2010, \aap, 511, A46

\bibitem[{{Miglio} {et~al.}(2009){Miglio}, {Montalb{\'a}n}, {Baudin},
  {Eggenberger}, {Noels}, {Hekker}, {De Ridder}, {Weiss}, \&
  {Baglin}}]{2009A&A...503L..21M}
{Miglio}, A., {et~al.} 2009, \aap, 503, L21

\bibitem[{{Miglio} {et~al.}(2010){Miglio}, {Montalb{\'a}n}, {Carrier}, {De
  Ridder}, {Mosser}, {Eggenberger}, {Scuflaire}, {Ventura}, {D'Antona},
  {Noels}, \& {Baglin}}]{2010A&A...520L...6M}
---. 2010, \aap, 520, L6

\bibitem[{Mihalas {et~al.}(1988)Mihalas, D\"appen, \& Hummer}]{mhd2}
Mihalas, D., D\"appen, W., \& Hummer, D.~G. 1988, ApJ, 331, 815

\bibitem[{{Monnier}(2003)}]{2003RPPh...66..789M}
{Monnier}, J.~D. 2003, Reports on Progress in Physics, 66, 789

\bibitem[{{Mosser}(2010)}]{2010AN....331..944M}
{Mosser}, B. 2010, Astronomische Nachrichten, 331, 944

\bibitem[{{Mosser} {et~al.}(2010){Mosser}, {Belkacem}, {Goupil}, {Miglio},
  {Morel}, {Barban}, {Baudin}, {Hekker}, {Samadi}, {De Ridder}, {Weiss},
  {Auvergne}, \& {Baglin}}]{2010A&A...517A..22M}
{Mosser}, B., {et~al.} 2010, \aap, 517, A22

\bibitem[{{Mosser} {et~al.}(2011{\natexlab{a}}){Mosser}, {Elsworth}, {Hekker},
  {Huber}, {Kallinger}, {Mathur}, {Belkacem}, {Goupil}, {Samadi}, {Barban},
  {Bedding}, {Chaplin}, {Garc\'\i a}, \& {et al.}}]{mosser2011}
---. 2011{\natexlab{a}}, \aap, submitted

\bibitem[{{Mosser} {et~al.}(2011{\natexlab{b}}){Mosser}, {Barban},
  {Montalb{\'a}n}, {Beck}, {Miglio}, {Belkacem}, {Goupil}, {Hekker}, {De
  Ridder}, {Dupret}, {Elsworth}, {Noels}, {Baudin}, {Michel}, {Samadi},
  {Auvergne}, {Baglin}, \& {Catala}}]{2011A&A...532A..86M}
---. 2011{\natexlab{b}}, \aap, 532, A86

\bibitem[{{Mosser} {et~al.}(2011{\natexlab{c}}){Mosser}, {Belkacem}, {Goupil},
  {Michel}, {Elsworth}, {Barban}, {Kallinger}, {Hekker}, {De Ridder}, {Samadi},
  {Baudin}, {Pinheiro}, {Auvergne}, {Baglin}, \&
  {Catala}}]{2011A&A...525L...9M}
---. 2011{\natexlab{c}}, \aap, 525, L9

\bibitem[{Nordlund(1982)}]{aake:numsim1}
Nordlund, {\AA}. 1982, A\&A, 107, 1

\bibitem[{Nordlund \& Stein(1991)}]{nordlundstein:RadDyn1991}
Nordlund, {\AA}., \& Stein, R.~F. 1991, in Stellar Atmospheres: Beyond
  Classical Models, ed. L.~C. et~al. (Dordrecht: Kluwer), 263--279

\bibitem[{{Nordlund} {et~al.}(2009){Nordlund}, {Stein}, \&
  {Asplund}}]{2009LRSP....6....2N}
{Nordlund}, {\AA}., {Stein}, R.~F., \& {Asplund}, M. 2009, Living Reviews in
  Solar Physics, 6, 2

\bibitem[{{Ohnaka} {et~al.}(2009){Ohnaka}, {Hofmann}, {Benisty}, {Chelli},
  {Driebe}, {Millour}, {Petrov}, {Schertl}, {Stee}, {Vakili}, \&
  {Weigelt}}]{2009A&A...503..183O}
{Ohnaka}, K., {et~al.} 2009, \aap, 503, 183

\bibitem[{{Ohnaka} {et~al.}(2011){Ohnaka}, {Weigelt}, {Millour}, {Hofmann},
  {Driebe}, {Schertl}, {Chelli}, {Massi}, {Petrov}, \&
  {Stee}}]{2011A&A...529A.163O}
---. 2011, \aap, 529, A163

\bibitem[{Prieto {et~al.}(2004)Prieto, Asplund, \&
  Bendicho}]{prieto:sun-line-LD}
Prieto, C.~A., Asplund, M., \& Bendicho, P.~F. 2004, ApJ, 423, 1109

\bibitem[{Prieto {et~al.}(2001)Prieto, Lambert, \&
  Asplund}]{prieto:SolarOI-abund}
Prieto, C.~A., Lambert, D.~L., \& Asplund, M. 2001, ApJ, 556, L63

\bibitem[{{Radau}(1880)}]{radau:quadratur}
{Radau}, R. 1880, J. Math. Pures et Appl., 6, 283

\bibitem[{Renka(1984)}]{renka:triangulation}
Renka, R.~J. 1984, ACM Trans. on Math. Softw., 10, 440

\bibitem[{{Robinson} {et~al.}(2004){Robinson}, {Demarque}, {Li}, {Sofia},
  {Kim}, {Chan}, \& {Guenther}}]{2004MNRAS.347.1208R}
{Robinson}, F.~J., {Demarque}, P., {Li}, L.~H., {Sofia}, S., {Kim}, Y., {Chan},
  K.~L., \& {Guenther}, D.~B. 2004, \mnras, 347, 1208

\bibitem[{Rosenthal {et~al.}(1999)Rosenthal, {Christensen-Dalsgaard}, Nordlund,
  Stein, \& Trampedach}]{rosenthal:conv-osc}
Rosenthal, C.~S., {Christensen-Dalsgaard}, J., Nordlund, {\AA}., Stein, R.~F.,
  \& Trampedach, R. 1999, A\&A, 351, 689

\bibitem[{{Schwarzschild}(1975)}]{1975ApJ...195..137S}
{Schwarzschild}, M. 1975, \apj, 195, 137

\bibitem[{{Stein} \& {Nordlund}(2000)}]{2000SoPh..192...91S}
{Stein}, R.~F., \& {Nordlund}, {\AA}. 2000, \solphys, 192, 91

\bibitem[{{Trampedach}(2010)}]{2010Ap&SS.328..213T}
{Trampedach}, R. 2010, \apss, 328, 213

\bibitem[{Trampedach {et~al.}(2011)Trampedach, {Christensen-Dalsgaard},
  Nordlund, Asplund, \& Stein}]{trampedach:T-tau}
Trampedach, R., {Christensen-Dalsgaard}, J., Nordlund, {\AA}., Asplund, M., \&
  Stein, R.~F. 2011, \aap, submitted

\bibitem[{Trampedach {et~al.}(1998)Trampedach, Christensen-Dalsgaard, Nordlund,
  \& Stein}]{trampedach:mons98}
Trampedach, R., Christensen-Dalsgaard, J., Nordlund, {\AA}., \& Stein, R.~F.
  1998, in Workshop on Science with a Small Space Telescope, ed. H.~Kjeldsen \&
  T.~R. Bedding (Denmark: Aarhus Universitet), 59--68

\bibitem[{{Unno}(1967)}]{1967PASJ...19..140U}
{Unno}, W. 1967, \pasj, 19, 140

\bibitem[{{V{\'a}zquez Rami{\'o}} {et~al.}(2005){V{\'a}zquez Rami{\'o}},
  {R{\'e}gulo}, \& {Roca Cort{\'e}s}}]{2005A&A...443L..11V}
{V{\'a}zquez Rami{\'o}}, H., {R{\'e}gulo}, C., \& {Roca Cort{\'e}s}, T. 2005,
  \aap, 443, L11

\end{thebibliography}

%\clearpage

\appendix

\section{Influence of the method and the initial guesses on the fitting}
\begin{figure}[htb]

\includegraphics[angle=90,width=9cm]{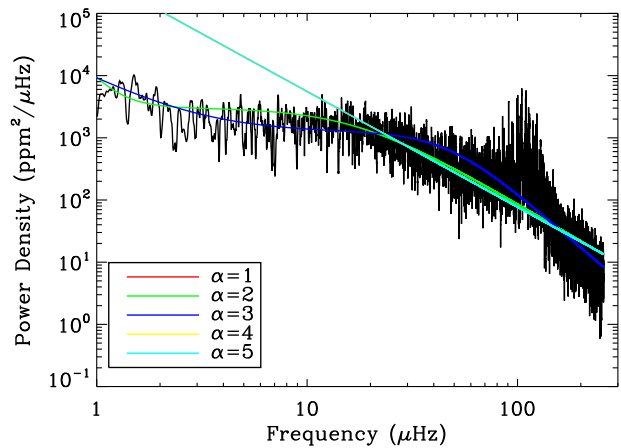}
\caption{PDS of KIC~11618103 smoothed over 10 bins with the comparison of the background fitting using 1 Harvey-like model and different values for $\alpha$. The fits using $\alpha$=1 and 4 are not visible because they are exactly the same as the fit using $\alpha$=5. \label{fig11}}
\end{figure}

We did a few tests with one of the methods used in this work (namely A2Z) to investigate the impact of the initial guess of the slope, $\alpha$, for the following values: [1,2,3,4,5] on six different red giants, ranging in $\nu_{\rm max}$ between 40 and 100~$\mu$Hz. To quantitatively compare the fits, we computed $\chi^2$ values as the mean value of the difference between the background fit and the heavily smoothed PDS. The best fits for the stars with $\nu_{\rm max} >$ 40~$\mu$Hz  were obtained for a slope of 2 and 3, with $\chi^2$ values ranging from 0.7 to 3.7, while for other values (1, 4, and 5), we could not reproduce the knee of the Harvey model and obtained only a straight line with a slope (see Figure~\ref{fig11}). For stars with $\nu_{\rm max} \sim$ 40~$\mu$Hz, the best fit is found for an initial guess of 3, 4, and 5 for the slope with $\chi^2$ values around 0.2.

We also noticed that by fixing the value of the white noise parameter the results were the same except in the case of a slope lower than 3. It seems that by assuming a slope too large and by adding a constrain on the white noise component, the background fit becomes less reliable. Then we added a Gaussian to the fit to incorporate the p-mode bump. These values were close to the ones obtained by not taking into account the bump of the oscillation modes.

We also checked how the fit converged when we decreased the initial guess of 1/$\tau_{\rm gran}$ to 10~$\mu$Hz. Though the code does not converge for high values of $\nu_{\rm max}$, the second branch starts to disappear for $\nu_{\rm max} < 40 \mu$Hz.

In the following, the fit done by A2Z with one Harvey-like model with an initial guess for 1/$\tau_{\rm gran}$ of 15~$\mu$Hz is chose as the reference fit.

We fitted two Harvey-like models using different initial values for time scale of the second Harvey model: 2, 5, 10 and 15~$\mu$Hz. Depending on the value of $\nu_{\rm max}$, the code could converge or not. For all the stars, a result for the fit could be obtained when the time scale value of the second model was 5 or 2~$\mu$Hz. For these stars, the $\chi^2$ between the fit and the PDS is smaller when the guess of the first slope is smaller than 3 and when we do not fix the value of the white noise parameter (around 3.6 compared to 7.5). We compared the dispersion of the values of $\tau_{\rm eff}$ and of $P_{\rm gran}$ with the uncertainties for the fits that reproduced the knee of the granulation and found that the dispersion is related to the way we fit the background. Depending on the initial guess of 1/$\tau_{\rm gran}$, we find that $\tau_{\rm eff}$ and $P_{\rm gran}$ vary by $\sim$~5\% around the value found by the reference fit. The uncertainties on these values are a $\sim$~6-10\% around the value found by the reference fit. We conclude that the dispersion of these parameters with the initial guess of 1/$\tau_{\rm gran}$ is of the same order of magnitude as the uncertainty. %is within the error of the fit. 

\begin{figure}[htb]

\includegraphics[angle=90,width=9cm]{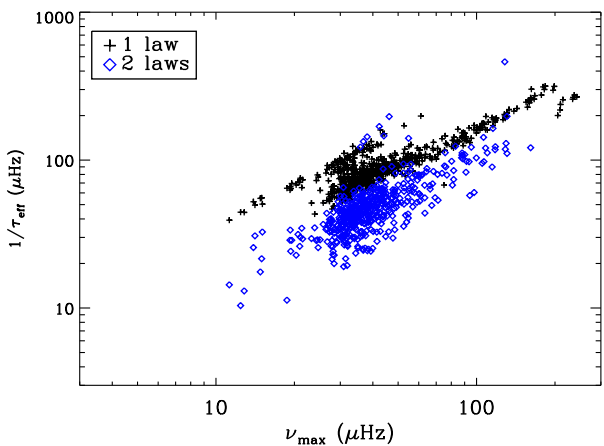}
\caption{Comparison of the background fitting using 1 Harvey-like model (black crosses) and 2 Harvey-like models (blue diamonds) for a subsample of red giants.\label{fig12}}
\end{figure}

We also performed some tests on 700 red giants by fitting two Harvey-like models without the white noise component and with a guess for the second $\tau$ of 300~$\mu$Hz. For these tests, a triangular smooth over 30 days was used to have enough points for the second Harvey like model. In Figure~\ref{fig12}, we can notice that for these cases, $\tau_{\rm eff}$ is larger compared to the case where we use only one Harvey model. As we fit two Harvey-like models, the code compensates the presence of the second component by increasing the value of $\tau_{\rm eff}$. Beside we find the same value of $a$ in the relation $\tau_{\rm eff}$=$(\nu_{\rm max})^{a}$ whether we fit one or two Harvey-like function while the second branch below 40~$\mu$Hz starts to disappear and to merge with the first one when we fit two Harvey-like functions.

Finally, we checked that the granulation signature in the PDS was not a result of observing some harmonics of the p-mode bump. We removed the p modes from the PDS by applying a pre-whitening and fitted the background with OCT. We obtained the same values for the granulation characteristics confirming that there was no reflection effect.

%% the file, uncomment the \plotone, \plottwo, and
%% If you need a layout that cannot be achieved with \plotone or
%% \plottwo, you can invoke the graphicx package directly with the
%% \includegraphics command or use \plotfiddle. For more information,
%% please see the tutorial on "Using Electronic Art with AASTeX" in the
%% documentation section at the AASTeX Web site,
%% http://www.journals.uchicago.edu/AAS/AASTeX.
%%
%% The examples below also include sample markup for submission of
%% supplemental electronic materials. As always, be sure to check
%% the instructions to authors for the journal you are submitting to
%% for specific submissions guidelines as they vary from

%% This example uses \plotone to include an EPS file scaled to
%% 80% of its natural size with \epsscale. Its caption
%% has been written to indicate that additional figure parts will be
%% available in the electronic journal.

\clearpage

\begin{deluxetable}{p{2cm}|ccc}
\tablecolumns{4}
\tabletypesize{\scriptsize}
%\rotate
\tablecaption{Summary of the background fitting methods (part 1).\label{tbl-1}}

\tablewidth{0pt}
\tablehead{
\colhead{ ID} & \colhead{ OCT}  & \colhead{ SYD} & \colhead{ CAN} 
}
%\tableline
\startdata
Model &  $W + \frac{\displaystyle 4\sigma^2\tau_{\rm gran}}{\displaystyle 1+(2\pi\nu\tau_{\rm gran})^\alpha}$  &  $W + \sum_{i=1}^2 \frac{\displaystyle 4\sigma_i^2 \tau_i} {\displaystyle 1 + (2\pi \nu \tau_i)^2 + (2\pi \nu \tau_i)^4}$  &  $W + \sum_{i=1}^3 \frac{\displaystyle 4\sigma_i^2\tau_i}{\displaystyle 1+(2\pi\nu\tau_i)^4}$   \\
 & \\
 Additional component & - & - & Gaussian function    \\
  & \\
Free parameters &  $\alpha$, $\sigma$, $\tau_{\rm gran}$, $W$ & $\sigma_i$, $\tau_i$, $W$ &  $\sigma_i$, $\tau_i$, $W$\\
 & \\
Minimization  &  Non-linear  & Levenberg-Markward  & Bayesian Markov-Chain  \\
 method & least squares  & least squares &  Monte Carlo \\
  & \\
Uncertainties &   Std dev. of & Std dev.  of & Posterior probability   \\
 &  parameters  & distributions from simulations & of density distribution 
 \enddata
%\tableline
%\end{tabular}
%% Any table notes must follow the \end{tabular} command.
%\tablenotetext{a}{Maximum a Posteriori}
%\end{center}
\end{deluxetable}

\begin{deluxetable}{p{2cm}|ccc}
\tablecolumns{4}
\tabletypesize{\scriptsize}
%\rotate
\tablecaption{Summary of the background fitting methods (part 2).\label{tbl-1_1}}

\tablewidth{0pt}
\tablehead{
\colhead{ ID} &  \colhead{ DLB} &  \colhead{ COR} & \colhead{ A2Z}
}
%\tableline
\startdata
Model &   $W + \frac{\displaystyle 4\sigma^2\tau_{\rm gran}}{\displaystyle 1+(2\pi\nu\tau_{\rm gran})^2}$ & $\frac{\displaystyle 4\sigma^2\tau_{\rm gran}}{\displaystyle 1+(2\pi\nu\tau_{\rm gran})^\alpha}$ & $W + \frac{\displaystyle 4\sigma^2\tau_{\rm gran}}{\displaystyle 1+(2\pi\nu\tau_{\rm gran})^\alpha}$  \\
 & \\
 Additional component &  - & - & Power law  \\
  & \\
Free parameters & $\sigma$, $\tau_{\rm gran}$, $W$ & $\alpha$, $\sigma$, $\tau_{\rm gran}$ & $\alpha$, $\sigma$, $\tau_{\rm gran}$, $W$\\
 & \\
Minimization  &    Linear  & Least squares & Maximum likelihood  \\
 method &  least squares &  & estimator \\
  & \\
Uncertainties &   -  & Std dev. of& Inversion of   \\
 &   &  parameters & hessian matrix
 \enddata
%\tableline
%\end{tabular}
%% Any table notes must follow the \end{tabular} command.
%\tablenotetext{a}{Maximum a Posteriori}
%\end{center}
\end{deluxetable}

\begin{table*}[htb]
\begin{center}
\caption{Values of the parameter $s$ in the power laws fitted in the Figures~\ref{fig1}, \ref{fig2}, and \ref{fig4}. The fits for OCT, A2Z, and all the methods together do not take into account the second branch. \label{tbl-2}}
\begin{tabular}{ccccc}
\tableline\tableline
Method &$\tau_{\rm eff} \propto (\nu_{\rm max})^{s}$ & $P_{\rm gran} \propto (\nu_{\rm max})^{s}$ &  $P_{\rm gran} \propto (\tau_{\rm eff})^{s}$   \\%$P_{\rm gran}/\tau_{\rm eff}$=$(log g)^{s}$ &
\tableline
OCT & -0.82~$\pm$~0.02& -1.2~$\pm$~0.02 &  2.11~$\pm$~0.03\\
SYD &-0.90~$\pm$~0.004 & -2.12~$\pm$~0.01 & 2.37~$\pm$~0.01 \\
CAN & -0.86~$\pm$~0.005 & -1.73~$\pm$~0.02 & 1.99~$\pm$~0.02 \\
DLB & -0.86~$\pm$~0.01 & -2.06~$\pm$~0.02 & 2.34~$\pm$~0.02 \\
COR & -0.90~$\pm$~0.005 & -2.15~$\pm$~0.12 &2.34~$\pm$~0.01\\
A2Z& -0.79~$\pm$~0.008 & -2.09~$\pm$~0.16 & 2.39~$\pm$~0.02 \\
\tableline
All & -0.89~$\pm$~0.005 & -1.90~$\pm$~0.01 &  2.19~$\pm$~0.01\\
%13 & 1070.58 $\pm$ 0.39\tablenotemark{b,c} & 1093.52 $\pm$0.45\tablenotemark{b,c} & 1064.34 $\pm$0.46\tablenotemark{b,c}\\
%14 & 1116.12 $\pm$ 0.81\tablenotemark{b,c} & ...& 1110.75 $\pm$0.53\tablenotemark{b,c}\\
\tableline
\end{tabular}
%% Any table notes must follow the \end{tabular} command.
\end{center}
\end{table*}

\end{document}